\documentclass[10pt,journal,cspaper,compsoc]{IEEEtran}
\pdfoutput=1

\ifCLASSOPTIONcompsoc
\else
   \usepackage{cite}
\fi
\usepackage{setspace}
\usepackage{graphicx}
\usepackage{subfig}
\usepackage[cmex10]{amsmath}

\usepackage{amsthm}
\usepackage{balance}

\ifCLASSINFOpdf
   \usepackage[]{graphicx}
\else
\fi
\usepackage[cmex10]{amsmath}

\usepackage{algorithm2e}
\usepackage{flushend}

\begin{document}
%\title{Bare Demo of IEEEtran.cls\\ for Computer Society Journals}

\title{Channel Selection Algorithm for Cognitive Radio Networks with Heavy-Tailed Idle Times}

\author{Senthilmurugan~S,~\IEEEmembership{Member,~IEEE}, Junaid~Ansari,~\IEEEmembership{Member,~IEEE}, Petri~M\"ah\"onen,~\IEEEmembership{Senior Member,~IEEE}, T.G.~Venkatesh, and Marina~Petrova,~\IEEEmembership{Member,~IEEE}

\IEEEcompsocitemizethanks{\IEEEcompsocthanksitem
Senthilmurugan S and T.G.~Venkatesh are with the Department of Electrical Engineering, Indian Institute of Technology Madras, Chennai, India. }
\IEEEcompsocitemizethanks{\IEEEcompsocthanksitem
Junaid Ansari, Petri~M\"ah\"onen and Marina~Petrova are with the Institute for Networked Systems at RWTH Aachen University, Germany.
 }
{\thanks{ The work has been supported in part by DFG (Deutsche Forschungsgemeinschaft) through UMIC research centre. The first author also acknowledges the funding of German Academic Exchange Service (DAAD) for his stay at RWTH Aachen University during the course of this work.}}

 }

\IEEEcompsoctitleabstractindextext{%
\begin{abstract}
We consider a multichannel Cognitive Radio Network (CRN), where secondary users sequentially sense channels for opportunistic spectrum access. In this scenario, the Channel Selection Algorithm (CSA) allows secondary users to find a vacant channel with the minimal number of channel switches. Most of the existing CSA literature assumes exponential ON-OFF time distribution for primary user’s (PU) channel occupancy pattern. This exponential assumption might be helpful to get performance bounds; but not useful to evaluate the performance of CSA under realistic conditions. An in-depth analysis of independent spectrum measurement traces reveals that wireless channels have typically heavy-tailed PU OFF times. In this paper, we propose an extension to the Predictive CSA framework and its generalization for heavy tailed PU OFF time distribution, which represents realistic scenarios. In particular, we calculate the probability of channel being idle for hyper-exponential OFF times to use in CSA. We implement our proposed CSA framework in a wireless test-bed and comprehensively evaluate its performance by recreating the realistic PU channel occupancy patterns. The proposed CSA shows significant reduction in channel switches and energy consumption as compared to Predictive CSA which always assumes exponential PU ON-OFF times.Through our work, we show the impact of the PU channel occupancy pattern on the performance of CSA in multichannel CRN.
\end{abstract}

}

\maketitle

\IEEEdisplaynotcompsoctitleabstractindextext
\IEEEpeerreviewmaketitle
\section{Introduction}
\IEEEPARstart{T}{he} recent spectrum measurement campaigns show that the fixed spectrum assignment policy for wireless devices results  in under utilization of allotted bandwidth  \cite{ Sadler2007, Liang2011, Wellens2009}. The scarcity of wireless spectrum has driven the researchers to design intelligent secondary network concepts to enable opportunistic access of spectrum free from licensed users (Primary Users). Towards this direction, the Dynamic Spectrum Access (DSA) approach has been proposed, in which the licensed bands are made available to unlicensed users (Secondary Users) while ensuring seamless operation for the licensed users \cite{Liang2011}. One of the most studied issues is the reuse of TV broadcasting bands, and more specifically TV White Spaces (TVWS). Two main regulators, FCC in the USA and Ofcom in the UK, have already issued regulations related to TVWS use in DSA fashion.

In parallel, many application-specific CRN MAC protocols which operate in frequency bands other than TVWS have been developed in literature. The design of CRN MAC protocols differs significantly from traditional MAC as they need to have close interaction between physical and MAC layer \cite{Cormio2009}. These application-specific CRN MAC protocols are optimized for a particular environment and work mostly in decentralized manner. In decentralized CRN MAC protocols, the secondary user has to sense the channel before transmission. This is due to the absence of centralized controller which allocates channels that are free from Primary Users (PU) to Secondary Users (SU). 

The major concern of multichannel CRN MAC protocols is that the sequential sensing of channels results in wastage of energy and time for SUs. In practical implementation, we also have to account for delays in  channel switching and TX/RX mode switching apart from channel sensing time. Thus, the secondary network throughput can be improved if SU uses Channel Selection Algorithm (CSA) that finds unoccupied channels by monitoring a smaller subset of channels based on \textit{a priori} probability of their idleness. In most of the existing literature, the channel occupancy pattern of PU is modelled either as two state Discrete Time Markov Chain (DTMC) or as exponential ON - OFF distribution \cite{Wellens2009}. The researchers have restricted themselves to the above models mostly due to analytical tractability and simplified  characterization of problems in CRN. These assumptions might not be valid in all scenarios. In fact, extensive spectrum measurement campaigns suggest heavy-tail distribution as an appropriate fit for PU channel occupancy pattern \cite{Wellens2009}. 

The impact of PU's channel activity pattern on spatial spectrum reuse opportunities has been studied in \cite{Riihij}. Macaluso \textit{et al.}  \cite{Macaluso2013} claims that the effectiveness of reinforcement learning algorithms for channel selection strongly depends on PU's channel activity pattern. In this paper, we design CSA framework where the channels are ordered based on their probability of being free from PU at a given time instant. The probability of particular channel being free from PU at a given time instant is a function of PU ON-OFF channel occupancy pattern and the previous sensing results (both sensing time and state) of that channel. We refer to the class of CSAs which predicts channel availability based on the above probability as Predictive CSA framework \cite{Kim2006}, \cite{Zhao2007}, \cite{Yang2008}. In this paper, we calculate the probability of channel being free from PU for a generalized PU ON-OFF time distribution, and in particular to Hyper-exponential Distribution (HED), which is a good  fit for heavy-tailed OFF times observed in spectrum measurement traces \cite{Lopez-Bentez2013}, \cite{Stabellini2010}. We then evaluate the performance of generalized predictive CSA framework using a large-scale USRP test-bed implementation \cite{Ettus} and a simple CRN MAC protocol for secondary network \cite{Labview}.

The rest of a paper is organized as follows: Section~\ref{sec:related} details related work in CSAs and the spectrum measurement trials. The system model and a brief overview of Predictive CSA framework are detailed in Section~\ref{system_model}. The Predictive CSA framework for generalized ON-OFF time distribution of primary user is proposed in Section~\ref{gen_predictive}. In Section~\ref{heavy_tail}, we derive the probability of idle channel for PU's heavy-tailed OFF time. The simplified structure of generalized predictive CSA framework for channels having i.i.d. PU ON-OFF time distributions is shown in Section~\ref{equivalence}. In Section~\ref{sec:MAC_design}, we discuss the simple multichannel CRN MAC protocol on which proposed CSA framework will be incorporated. Then, we evaluate the performance of generalized predictive CSA framework in realistic scenario with  wireless test-bed in Section~\ref{experiment} followed by conclusion in Section~\ref{conclusion}.

\section{Related work} \label{sec:related}
CRN MAC protocols differ significantly from traditional MAC protocols as SU can only use channels that are free from PU transmissions. In scenarios where the secondary user has to sequentially sense channels, the CSA plays a vital role in finding vacant channels with minimal number of channel switches. The performance of these CSAs inherently depend on the traffic distribution of PUs. In this section, we review the related work in CSAs and spectrum measurements.
 \begin{figure*}[ht]
\centering  
 \includegraphics[scale=0.7, trim=0.95cm 0.5cm 0mm 0.5cm,clip=true]{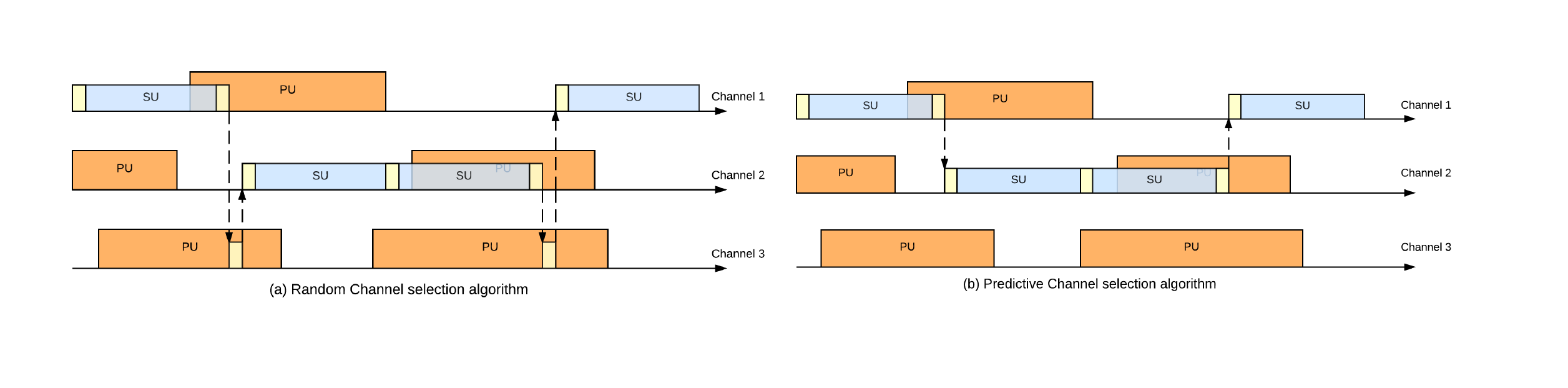} 
\caption{\small The snapshot of activities in multi-channel CRN if the secondary transmitter uses (a) Randomly selects a channel for sensing (b) Intelligently selects a channel for sensing.}
\label{fig:inteligent_CSA}
\end{figure*}
\subsection{Channel Selection Algorithm}
The channel selection algorithms for multichannel  CRN MAC protocols can be roughly classified into (i) learning based CSAs and (ii) ON-OFF distribution based CSAs depending on whether the statistics of a channel is known \textit{a priori} or estimated on run-time. In learning based CSAs, the statistics of PU's channel usage are not known to the secondary user and thus they have to learn these parameters \cite{Song2007}. Since it is very difficult to learn the ON-OFF time distribution of PU at run-time, these algorithms rely on the first and second order statistics of ON-OFF time distributions \cite{Song2007,Auer2002}.  In ON-OFF distribution based CSAs, the ON-OFF time distribution of PUs are used and they tend to be more efficient than learning based CSAs. Kim \textit{et al.} \cite{Kim2006} have calculated the probability of channel being idle in future for the exponential ON-OFF time distribution using the knowledge of previous sensing result (both sensing time instant and result). The calculated probability is used in ordering the channels for sensing in multichannel cognitive networks. Yang \textit{et al.} \cite{Yang2008} have modified the above algorithm for PUs with fixed ON - exponential OFF time distribution. We refer the above class of algorithms \cite{Kim2006}, \cite{Yang2008} as Predictive CSAs. The Partially Observable Markov Decision Process (POMDP) based \textit{Greedy CSA} framework has been proposed by Zhao \textit{et al.} \cite{Zhao2007} where the PU channel occupancy is modelled as discrete time ON-OFF Markov chain.  This framework has been extended to unslotted PU with exponential ON-OFF time distribution in \cite{Zhao2008a}. Most of these CSAs in literature \cite{Kim2006}, \cite{Zhao2008a} assume exponential idle time distribution which is not realistic for channels with low PU duty cycle \cite{Wellens2009}. 

\subsection{Spectrum measurements}\label{sec:measurements}
 The primary user activity pattern on cellular band has been characterized in \cite{Willkomm2009} where the authors report log-normal type of distributions for the call duration. Wellens \textit{et al.} \cite{Wellens2009} approximate the idle times of low traffic channel and the busy time of heavy traffic channels with heavy-tailed distributions. The authors of paper \cite{Wang2012} have also discussed the origins of heavy-tailed distribution in secondary network based on the observation of PU's ON times in \cite{Wellens2009}. Miguel \textit{et al.} \cite{Lopez-Bentez2013} suggest separate distribution fits for low and high time-resolution. The low-time resolution of idle times are fitted with generalized Pareto distribution where as the distribution of high-time resolution idle times depends on the technology deployed in the spectrum band \cite{Lopez-benitez2011}. 
 
The authors of \cite{Geirhofer2006} model 802.11 MAC behaviour as continuous-time semi-Markov process and analytically derive generalized Pareto distribution for channel idle times in a controlled environment with only WLANs. Stabellini \cite{Stabellini2010} considers heterogeneous traffic from WLANs/Bluetooth on 2.4 GHz band and fits spectrum traces with hyper-exponential distribution (HED) for idle times. Liu \textit{et al.} \cite{Liu2012} shows that the Complementary Cumulative Distribution Function (CCDF) of channel idle time has power law decay  upto some critical time after which it has exponential decay. The data set with above behaviour can be well modelled with hyper-exponential distributions (HED) \cite{Karagiannis2007}.  We have also found that the CCDF of idle times of Wellens \textit{et al.} \cite{Wellens2009} GSM traces have power-law decay with exponential tail. The above observations motivate us to design CSA for channels exhibiting heavy-tailed idle times.

\subsection{Major Contributions of this Work}
The main contributions of this paper that differentiate our work from the existing literature are as follows:
\begin{itemize}
\item Although there are many proposals of CSAs and various spectrum measurement campaigns, only few works in literature have used realistic distribution while designing MAC layer algorithms such as CSA. We use the knowledge obtained from spectrum measurement campaigns to enable efficient MAC layer algorithms and radio resource management.  
\item Secondly, our paper is one among very few works which has actually implement CSA on a large-scale wireless test-bed.  The impact of primary user's idle time distribution on the CSA is evaluated by recreating realistic PU channel occupancy patterns over-the-air on our wireless test-bed.
\item In the process of deducing simplified structure of CSA in i.i.d. PU channels, we have proven the equivalence of predictive and POMDP-based greedy CSA \cite{Zhao2007}, which were treated until now as separate lines of work. 
\end{itemize}

Our work is towards the direction of using the knowledge obtained in spectrum measurement
campaigns in efficient design of MAC layer algorithms. This is a unique angle to this work and we
believe that our work provides significant insights into realistic scenarios with comprehensive over-
the-air evaluation.

\section{System model and CSA Overview} \label{system_model}
\subsection{System model}
We consider $N_{c}$ channels each being licensed to a primary user.  The channel usage time (ON time) and inter-arrival time (OFF time) of primary user in a channel $a$ is characterized by probability distribution functions $f_{X}^{a}$ and $f_{Y}^{a}$, respectively.  The ON-OFF time distribution of PUs in these $N_{c}$ channels are assumed to have independent and non-identical distribution. This assumption was validated by Ghosh \textit{et al.} \cite{Ghosh2012} using the measurement traces of Wellens \textit{et al} \cite{Wellens2009}. Without loss of generality, we assume that the secondary network consists of single transmitter-receiver pair that opportunistically access one of these $N_{c}$ channels. In our work, we assume that the secondary transmitter sequentially sense the channels until a free channel is detected. Let $t_{sense}$ be secondary transmitter's channel sensing duration needed to reliably estimate channel occupancy. If the channel is sensed idle, the secondary transmitter uses the channel for  $\Delta t$ duration. If the channel sensing result is busy, the secondary transmitter selects a new channel according to CSA and repeats the sensing process.

When the secondary transmitter efficiently selects a new channel for sensing, it reduces the channel switch cost and the time spent in finding the idle channel. We have chosen Predictive CSA from \cite{Kim2006} as basic framework for our work. The Predictive CSA in \cite{Kim2006} considers only the exponential ON-OFF time distributions of primary user. In our work, we extend this framework for generalized ON-OFF time distribution, and in particular to heavy-tailed OFF time distribution.

\subsection{Brief overview of Predictive CSA framework }

Let us assume that the first channel sensed by the SU is occupied by PU as shown in Fig.~\ref{fig:inteligent_CSA}. As a result, SU has to move to a new channel for sensing. If the secondary transmitter intelligently selects a new channel for sensing, it yields a lesser number of channel switches while finding an idle channel. In Predictive CSA framework, the secondary transmitter maintains a belief vector, $\Lambda^{pr}(t)$, of dimension $N_{c}$ whose elements are the probability of corresponding channel being free from primary user at time $t$. Let $\omega^{pr}_{a}(t)$ be the probability of channel $a$ being free from PU and is calculated based on the ON-OFF time distribution of PU that is currently using channel $a$ and the last sensing result. Hence, the belief vector $\Lambda^{pr}$ is 
\begin{equation}
\Lambda^{pr}(t) = [\omega_{1}^{pr}(t) ,\omega_{2}^{pr}(t) \ldots ,\omega_{Nc}^{pr}(t)]. \nonumber
\end{equation}
Then, the secondary transmitter selects a channel $a^{*}$ which has the maximum probability of being free from primary user, i.e.
\begin{equation}
a^{*}(t) = \arg\lbrace\max_{a=1,..,N_{c}} \omega_{a}^{pr}(t) \rbrace. \nonumber
\end{equation}

If the channel $a^{*}$ is sensed busy, the secondary user selects a channel which has the next maximal probability of being free from PU. Further, it also updates the probability $\omega_{a^*}^{pr}$. If the newly  selected channel is sensed free from PU, the secondary transmitter uses the channel for $\Delta t$ duration. To limit the interference to PU, the secondary user has to sense the channel for every $\Delta t$ duration. When PU occupies the channel, the secondary network will know about it in the next sensing instant and shifts to a new channel based on predictive CSA framework.

\section{Generalized Predictive CSA framework} \label{gen_predictive}
\begin{figure}[ht]
\centering
 \includegraphics[scale=0.48, trim=7mm 10mm 0mm 15mm,clip=true]{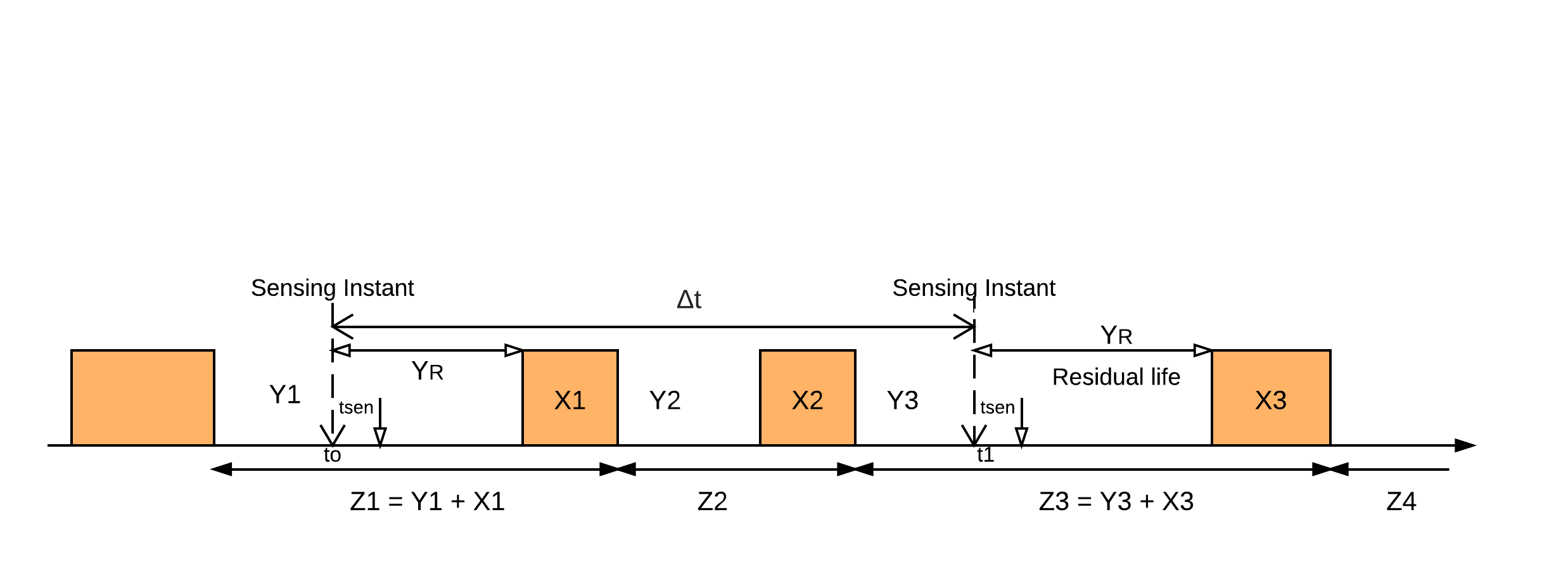} 
\caption{ The channel activity of a primary user and a secondary user on a particular channel.}
\label{fig:PU_activity}
\end{figure}

In this section, we will derive the probability of channel being free from PU $\omega^{pr}(t)$ given the ON-OFF time distributions of PU and the last sensing result and sensing time. For simplicity, let us consider one particular channel for derivation of probability $\omega^{pr}(t)$. We note that this approach can be then straightforwardly applied to all other channels. A typical case of the PU's activity on particular channel is shown in Fig.~\ref{fig:PU_activity}. Let $\{X_{i}\}$ and $\{Y_{i}\}$ be the random variables denoting the ON  and the OFF times of PU on the channel, respectively. The probability density function of the ON and the OFF times of PU are given as $f_{X}$ and $f_{Y}$ and are assumed to be statistically independent. Without loss of generality, let us assume the process starts with PU being in OFF state with holding time $Y_{1}$ followed by ON state with the holding time $X_{1}$. We will call this ON-OFF cycle as the renewal cycle with $Z_{i} = Y_{i} + X_{i}$ as $Y_{i}$'s and $X_{i}$'s are statistically independent of each other. Hence, the distribution of $Z$ is given by the convolution of the ON time and the OFF time distribution of PU, i.e. $f_{Z} = f_{Y} \ast f_{X}$. The Laplace transform of $f_{Z}$ is written as $f_{Z}^{*}(s) = f_{Y}^{*}(s) f_{X}^{*}(s)$. 

\subsection{Probability of Idle channel given the last sensing instant falls in PU OFF time} \label{last_off}
Suppose that SU's first sensing instant falls in OFF time of primary user. The secondary user uses the channel for $\Delta t$ duration. The next sensing instant of secondary transmitter falls at time $t_1 = t_0+\Delta t$ as shown in Fig 2. For the channel to be free from primary user at time $t_1$, the primary user activity in channel should be either of following cases:
\begin{enumerate}
\item[(i)] There is no arrival of the PU in the interval $[t_0,t_1]$.
\item[(ii)] The primary user might have used the channel at least once in $(t_0,t_1)$ but it would have left the channel by time $t_{1}$.
\end{enumerate}

Thus the probability of channel being free from PU at time $t_1$ given the last sensing instant $t_0$ falls in PU OFF time, denoted as $P_{OFF,OFF}([t_{0},t_{1}])$, is 
\begin{equation}
P_{OFF,OFF}([t_{0},t_{1}]) = Pr(\textit{Case {i}}) + Pr(\textit{Case {ii}}).
\end{equation} 

\subsubsection*{\textit{Case (i): No PU arrival in time interval $[t_0,t_1]$ }}
The scenario of no PU arrival in $[t_{0},t_{1}]$ is equivalent to saying that the remaining OFF time of primary user exceeds time $t_{1}$. Thus, the probability of no PU arrival in $[t_{0},t_{1}]$ is equal to the probability of remaining OFF time being greater than $(t_{1}-t_{0})$, i.e. we have  
\begin{equation}
\begin{split}
Pr(\textit{No PU arrival in }[t_{0},t_{1}]) & = Pr( Y_{R} \geq t_{1}-t_{0} ) 
\\& = Pr(Y_{R} \geq \Delta t ).  \nonumber
\end{split}
\end{equation}
The remaining OFF time $Y_{R}$ has probability density function \cite{Cox}
\begin{equation}
f_{Y_{R}}(r) = \frac{1}{E(Y)}  Pr(Y \geq r). \nonumber
\end{equation}
Thus, the probability of no primary user arriving  in $[t_{0},t_{1}]$ is given by
	\begin{equation}
	\begin{split}
	Pr(\textit{No PU arrival in }&[t_{0},t_{1}]) = Pr(Y_{R} \geq \Delta t) \\
	& = \frac{1}{E(Y)} \int_{\Delta t}^{\infty} Pr(Y \geq u)du.
	\end{split}
   \end{equation}

\subsubsection*{\textit{Case (ii): PU uses channel at least once in $[t_{0},t_{1}]$ but vacates the channel before $t_{1}$}}
Let the random variable $S_{n}$ denote the time for $n^{th}$ OFF-ON renewal cycle of PU from the previous sensing instant $t_{0}$. 
\begin{equation}
S_{n} = \tilde{Z_{1}}  + Z_{2} + \ldots + Z_{n}, \nonumber
\end{equation} 
where $Z_{k}$ has the distribution $f_{Z} = f_{Y}*f_{X} \text{ for } k=2,3, \dots n$.
However, the first renewal cycle $\tilde{Z_{1}}$ has different distribution from $Z_{k}$ as the residual OFF time has to be used in calculation of $f_{\tilde{Z_{1}}}$ and thus $\tilde{Z_{1}}$ has distribution $f_{\tilde{Z_{1}}} = f_{Y_{R}}*f_{X}$. Since the random variables $\tilde{Z_{1}}$ and $Z_{k}$ are independent\footnote[1]{Since $Y_1$ and $X_1$ are independent (system model), $Y_R$ and $X_1$ are also independent. Secondly, we note that $X_1$, $Y_R$, $X_k$, $Y_k$ are all independent of each other. Hence, $\tilde{Z}_1 = Y_R + X_1$ and $Z_k = Y_k + X_k$ are also independent. }, the distribution of $S_{n}$ is given as
\begin{equation}
 f_{S_{n}} = f_{\tilde{Z_{1}}} *  {f_{Z}}^{(n-1)}, \nonumber
\end{equation}
where $f_{Z}^{(n-1)}$ is the (n-1) fold convolution of $f_{Z}$ with itself.
The probability of PU using the channel at least once in $[t_{0},t_{1}]$ but vacates before the next sensing instant is calculated as follows: 
\begin{equation}
\begin{split}
Pr(\textit{PU at least once}) =  &  \sum_{i=1}^{\infty} \int_{0}^{\Delta t} f_{S_{i}}(u)P(Y > \Delta t - u)du 
\end{split}, 
\end{equation}
where the index $i$ denotes the number of renewals in the interval $[t_{0},t_{1}]$.

Thus the conditional probability of channel being idle given the last sensing instant falls in PU OFF time is
\begin{equation}
\begin{split}
P_{OFF,OFF}([t_{0},t_{1}])& =  \frac{1}{E(Y)} \int_{\Delta t}^{\infty} Pr(Y \geq u)du \\ 
& + \sum_{i=1}^{\infty} \int_{0}^{\Delta t}  f_{S_{i}}(u)P(Y > \Delta t - u)du. 
\end{split} 
\end{equation}
The value of $P_{OFF,OFF}([t_{0},t_{1}])$ depends only on the duration $\Delta t$ and hence, it can be written as $P_{OFF,OFF}(\Delta t)$. The complexity involved in the calculation of  $f_{S_{i}} = f_{\tilde{Z_{1}}} *  {f_{Z}}^{(i-1)}$, which requires $i$-fold convolution, is very high in above equation. However, the $i$-fold convolution in time domain turns to  $i$-fold multiplications in Laplace domain $f_{S_{i}}^{*}$. The Laplace transform of $P_{OFF,OFF}(\Delta t)$ in Eq.~(4) is written as,
\begin{equation}
P_{OFF,OFF}^{*}(s) = \frac{1}{s} - \frac{(1-f_{Y}^{*})(1-f_{X}^{*})}{s^2 E(Y)(1-f_{X}^{*}f_{Y}^{*})},
\label{eq:p11} 
\end{equation}
where $f^{*}_{X}$ and $f^{*}_{Y}$ are the Laplace transform of $f_X$ and $f_Y$, respectively. The inverse Laplace transform of $P_{OFF,OFF}^{*}(s)$, after substituting $f_{X}^{*}$ and $f_{Y}^{*}$ in Eqn(4), yields the required probability $P_{OFF,OFF}(\Delta t)$.

\subsection{Probability of Idle channel given the last sensing instant falls in PU ON time}

The probability of channel being free from PU given the last sensing instant falls in PU ON time, denoted as  $P_{ON,OFF}(\Delta t)$, can be calculated in similar manner as in Section~\ref{last_off}. However, we follow a simple approach to calculate $P_{ON,OFF}(\Delta t)$ , first by deriving $P_{ON,ON}(\Delta t)$, as follows: Let $P_{ON,ON}$ denote the probability of channel being used by PU given the last sensing instant falls in PU ON time. Then, the $P_{ON,ON}^{*}(s)$ is calculated by interchanging the role of $f_{X}$ and $f_{Y}$ in $P_{OFF,OFF}^{*}(s)$.
\begin{equation}
P_{ON,ON}^{*}(s) = \frac{1}{s} - \frac{(1-f_{Y}^{*})(1-f_{X}^{*})}{s^{2}E(X)(1-f_{X}^{*}f_{Y}^{*})} .\nonumber
\label{eq:p00}
\end{equation}
Thus, the required probability $P_{ON,OFF}(\Delta t)$ is obtained by taking the inverse Laplace transform of $P_{ON,OFF}^{*}(s)$ which is given by
\begin{equation}
P_{ON,OFF}^{*}(s) = \frac{1}{s}- P_{ON,ON}^{*}(s).
\end{equation}\label{eqn:p10}

Now, we will define the probability of channel being free from primary user, $\omega^{pr}(t)$, used in Predictive CSA framework as
\begin{equation}
\begin{split}
\omega^{pr}(t) & = \begin{cases}
P_{OFF,OFF}(\Delta t),  &\textit{if PU OFF in last sensing}\\ 
P_{ON,OFF}(\Delta t),  &\textit{if PU ON in last sensing}, 
\end{cases}
\end{split}
\end{equation}
where $\Delta t$ is the time duration between current time $t$ and last sensing time. In general, the probability $\omega_{a}^{pr}(t)$ for any channel $a$ is given as
\begin{equation}
\begin{split}
\omega^{pr}_{a}(t) & = \begin{cases}
P_{OFF,OFF}^{a}(\Delta t_{a}),  &\textit{if PU OFF in last sensing}\\ 
P^{a}_{ON,OFF}(\Delta t_{a}),  &\textit{if PU ON in last sensing},
\end{cases} \nonumber
\end{split}
\end{equation}
where $P^{a}_{ON,OFF}$ and $P_{OFF,OFF}^{a}$ are the probabilities of channel $a$ being free from PU given $f^{a}_X $ \& $f^{a}_Y$ and previous sensing results, and $\Delta t_{a}$ is the time duration between current time $t$ and the last sensing instant of the channel $a$.

\begin{figure*}[ht]
\begin{minipage}[b]{0.5\linewidth}
\centering 
\includegraphics[scale=0.45, trim=7mm 0mm 0mm 0mm,clip=true]{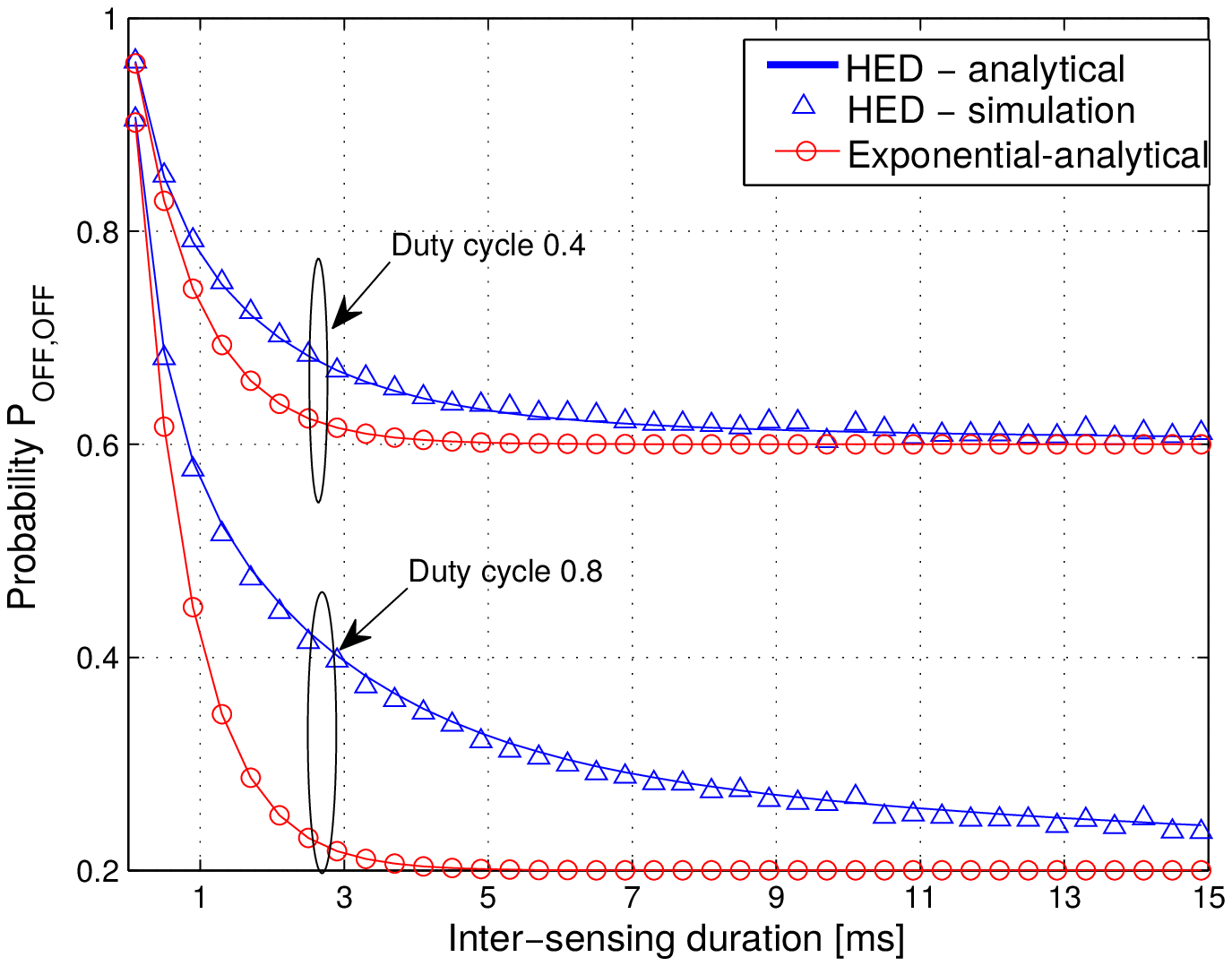} 
\caption{\small Comparison of probability  $P_{OFF,OFF}$ for various OFF time distributions (Exponential \& HED) versus sensing interval $\Delta t$ for different PU duty cycles  \cite{Stabellini2010}).} 
\label{fig:p11_04} 
\end{minipage}
\hspace{0.5cm}
\begin{minipage}[b]{0.5\linewidth}
\centering 
 \includegraphics[scale=0.45, trim=7mm 0mm 0mm 0mm,clip=true]{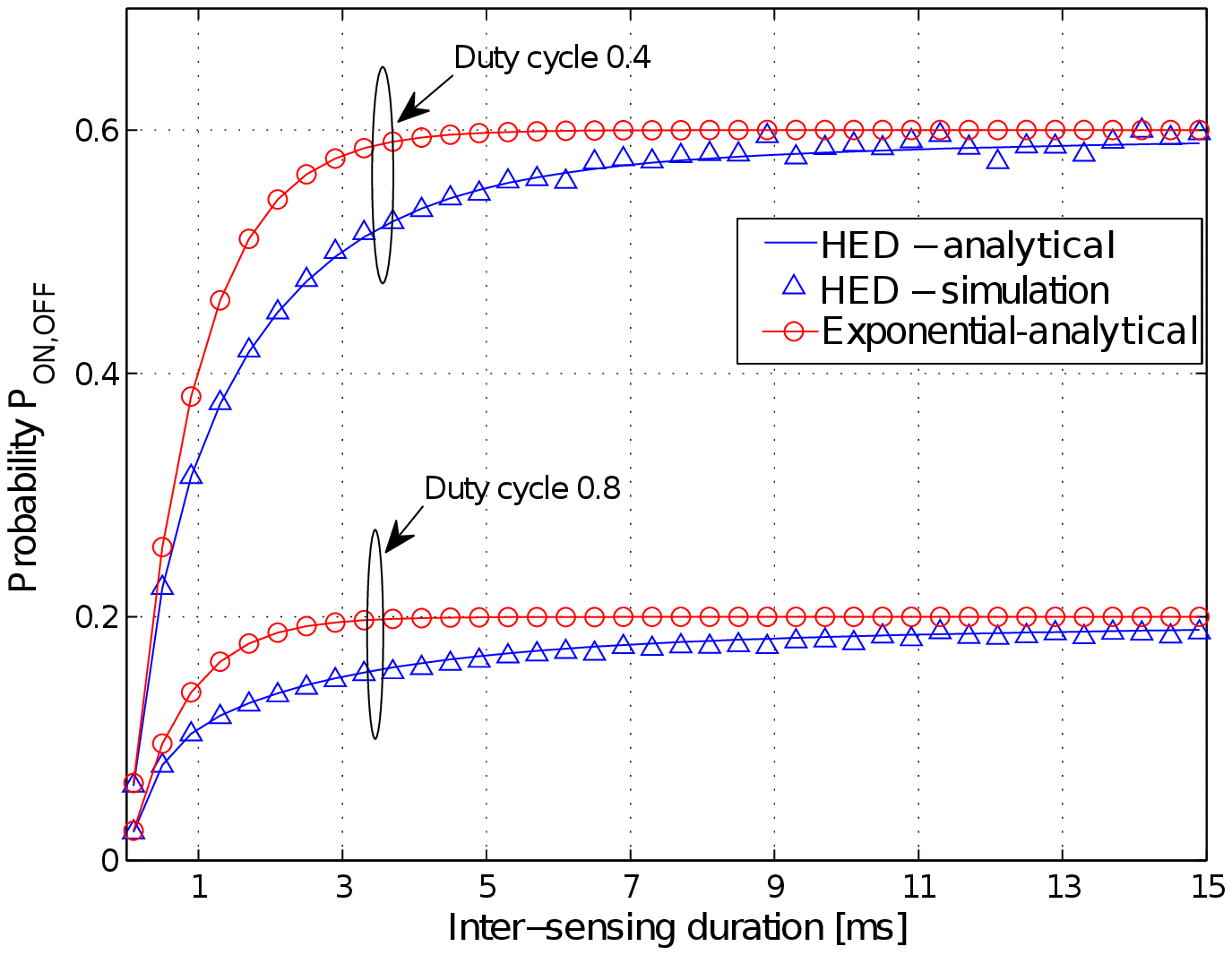} 
\caption{\small Comparison of  probability  $P_{ON,OFF}$ for various OFF time distributions (Exponential \& HED ) versus sensing interval $\Delta t$ for different PU duty cycles (HED parameters in \cite{Stabellini2010}). }
\label{fig:p01_04}
\end{minipage}
\end{figure*}

\section{probability of idle channel for heavy-tailed idle times} \label{heavy_tail}

The spectrum measurement campaigns in Section~\ref{sec:measurements} suggest heavy-tailed idle time distributions such as generalized Pareto and log-normal for frequency bands with light traffic or low occupancies. But these distributions do not have closed-form expression for Laplace transform which is required for derivation of $P_{OFF,OFF}$ and $P_{ON,OFF}$. For ease of analytical modelling, most of the existing  CRN literature assume exponential ON-OFF times for PU. These assumptions might be helpful in getting bounds but not for deriving CSA algorithm that inherently depends on realistic PU ON-OFF time distributions. In this section, we derive the probability of channel being idle for a heavy-tailed PU OFF times to use in CSA framework.

\subsection{Approximation of heavy-tailed idle times}
At this juncture, we apply methods from queuing theory literature to approximate the heavy-tailed distributions with other simpler distributions. Feldmann \textit{et al.} \cite{Feldmann1997} have proposed a method to fit mixtures of exponential to heavy-tailed distributions that are monotonically decreasing functions. Similarly, non-monotonic heavy tailed function like log-normal can be fitted with generalized hyper-exponential distribution \cite{Yu2012}. Dufresne \cite{Dufresne2007} has provided methods to fit any probability distribution function by Hyper-Exponential distribution (HED) with a small number of phases. The readers are directed to \cite{Dufresne2007} for more details on distribution approximations. Recent study on spectrum measurement traces \cite{Liu2012}, \cite{Sharma2011} have also proposed HED based channel occupancy model. In our work, we assume that a very good HED fit is already  available for a channel idle times for deriving the required probabilities and use them in generalized predictive CSA framework.

\subsection{Probability of Idle channel for HED OFF times}
 Let the p.d.f. of HED fitted for a heavy-tailed PU OFF times be
\begin{equation}
 f_{Y}=\sum_{i=1}^{N} p_{i} \lambda_{i} e^{-\lambda_{i} t}, \nonumber
\end{equation} 
where $\sum_{i=1}^{N} p_{i} = 1$
 and the exponential ON time distribution be  $ f_{X} = \lambda_{ON} e^{-\lambda_{ON} t} $  . The Laplace transform of these distributions are
 \begin{equation}
  f_{X}^{*}(s) = \frac{\lambda_{ON}} {s + \lambda_{ON}}, \quad
  f_{Y}^{*}(s)=\sum_{i=1}^{N} \frac{p_{i} \lambda_{i}}{s + \lambda_{i}}. \nonumber
  \end{equation}
 Substituting the above $f_X^*(s)$ and $f_Y^*(s)$ in Eq.~(\ref{eq:p11}), we get 
\begin{equation}
P_{OFF,OFF}^{*}(s) = \frac{\sum_{i=1}^{N} \frac{p_{i}}{s+\lambda_{i}}}{s+\lambda_{ON}\sum_{i=1}^{N} \frac{p_{i}}{s+\lambda_{i}}}. \nonumber
\label{eq:pffs}
\end{equation}
 
The inverse Laplace transform is found by factorizing the above equation in to partial fraction form as follows:  We first rearrange the equation of $P_{OFF,OFF}^*(s)$ as,
\begin{equation}
P_{OFF,OFF}^{*}(s) = \frac{1}{s} - \frac{1}{E(Y)} \frac{N(s)}{sD(s)},
\label{eq:form}
\end{equation}
where $N(s)$ and $D(s)$ are the numerator and the denominator polynomials in $s$-domain. As the degree of $N(S)$ is less than $sD(S)$, we use the \textit{residue theorem} \cite{Cheney} to find the inverse Laplace transform as
\begin{equation}
\begin{split}
P_{OFF,OFF}(\Delta t) = 1-\frac{1}{E(Y)}( \lim_{s\to0} \frac{ sN(s)e^{s\Delta t}}{s\prod_{k=1}^{N}(s-r_{k})} \\ + \sum_{i=1}^{N} \lim_{s\to r_{i}} \frac{ (s-r_{i})N(s)e^{s\Delta t}}{s\prod_{k=1}^{N}(s-r_{k})}),
\end{split}
\end{equation}
\begin{equation}
\begin{split}
P_{ON,ON}(\Delta t) = 1-\frac{1}{E(X)}( \lim_{s\to0} \frac{ sN(s)e^{s\Delta t}}{s\prod_{k=1}^{N}(s-r_{k})} \\ + \sum_{i=1}^{N} \lim_{s\to r_{i}} \frac{ (s-r_{i})N(s)e^{s\Delta t}}{s\prod_{k=1}^{N}(s-r_{k})}),
\end{split}
\end{equation}
where $\{r_{k}\}$ are the roots of denominator polynomial $D(S)$ in Eq.~(\ref{eq:form}). Studies on spectrum measurement traces have shown that the channel idle times can be well-approximated by HED with less number of phases (around 3 $\sim$ 4) \cite{Liu2012}. The probabilities $P_{OFF,OFF}(\Delta t)$ and $P_{ON,ON}(\Delta t)$ for exponential ON -- 3 phase HED OFF times are given in Eqs.~(\ref{eq:floatingequation}) and (\ref{eq:floatingequation2}), respectively.
\newcounter{tempequationcounter}
\begin{figure*}[!t]
\normalsize
\setcounter{tempequationcounter}{\value{equation}}
\begin{IEEEeqnarray}{rCl}
\setcounter{equation}{11}
P_{OFF,OFF}(\Delta t) = \begin{cases} 1-\frac{1}{E(Y)}( \frac{p_{1}\lambda_{2}\lambda{3} + p_{2}\lambda_{1}\lambda_{3} + p_{3}\lambda_{1}\lambda_{2}}{(-r_{1}r_{2}r_{3})} 
 + \frac{p_{1}(r_{1}+\lambda_{2})(r_{1}+\lambda_{3}) + p_{2}(r_{1}+\lambda_{1})(r_{1}+\lambda_{3}) + p_{3}(r_{1}+\lambda_{1})(r_{1}+\lambda_{2}) }{r_{1}(r_{1}-r_{2})(r_{1}-r_{3})} e^{r_{1}\Delta t} \\
 + \frac{p_{1}(r_{2}+\lambda_{2})(r_{2}+\lambda_{3}) + p_{2}(r_{2}+\lambda_{1})(r_{2}+\lambda_{3}) + p_{3}(r_{2}+\lambda_{1})(r_{2}+\lambda_{2}) }{r_{2}(r_{2}-r_{1})(r_{2}-r_{3})}e^{r_{2}\Delta t} \\
 + \frac{p_{1}(r_{3}+\lambda_{2})(r_{3}+\lambda_{3}) + p_{2}(r_{3}+\lambda_{1})(r_{3}+\lambda_{3}) + p_{3}(r_{3}+\lambda_{1})(r_{3}+\lambda_{2}) }{r_{3}(r_{3}-r_{1})(r_{3}-r_{2})}e^{r_{3}\Delta t}). 
\end{cases}
\label{eq:floatingequation}
\end{IEEEeqnarray}
\setcounter{equation}{\value{tempequationcounter}}
\hrulefill
\vspace*{4pt}
\end{figure*}
\begin{figure*}[!t]
\normalsize
\setcounter{tempequationcounter}{\value{equation}}
\begin{IEEEeqnarray}{rCl}
\setcounter{equation}{12}
P_{ON,ON}(\Delta t) = \begin{cases} 1-\frac{1}{E(X)}( \frac{p_{1}\lambda_{2}\lambda{3} + p_{2}\lambda_{1}\lambda_{3} + p_{3}\lambda_{1}\lambda_{2}}{(-r_{1}r_{2}r_{3})} 
 + \frac{p_{1}(r_{1}+\lambda_{2})(r_{1}+\lambda_{3}) + p_{2}(r_{1}+\lambda_{1})(r_{1}+\lambda_{3}) + p_{3}(r_{1}+\lambda_{1})(r_{1}+\lambda_{2}) }{r_{1}(r_{1}-r_{2})(r_{1}-r_{3})} e^{r_{1}\Delta t} \\
 + \frac{p_{1}(r_{2}+\lambda_{2})(r_{2}+\lambda_{3}) + p_{2}(r_{2}+\lambda_{1})(r_{2}+\lambda_{3}) + p_{3}(r_{2}+\lambda_{1})(r_{2}+\lambda_{2}) }{r_{2}(r_{2}-r_{1})(r_{2}-r_{3})}e^{r_{2}\Delta t} \\
 + \frac{p_{1}(r_{3}+\lambda_{2})(r_{3}+\lambda_{3}) + p_{2}(r_{3}+\lambda_{1})(r_{3}+\lambda_{3}) + p_{3}(r_{3}+\lambda_{1})(r_{3}+\lambda_{2}) }{r_{3}(r_{3}-r_{1})(r_{3}-r_{2})}e^{r_{3}\Delta t}).
\end{cases}
\label{eq:floatingequation2}
\end{IEEEeqnarray}
\setcounter{equation}{\value{tempequationcounter}}
\hrulefill
%\vspace*{4pt}
\end{figure*}
\addtocounter{equation}{2} 

\subsection{\textit{Comparison of probabilities for exponential  and HED OFF times}}

When the number of phases of HED OFF time is set as $N = 1$, the probabilities $P_{OFF,OFF}(\Delta t)$ and $P_{ON,OFF}(\Delta t)$ for exponential PU ON-OFF distribution with parameters $\lambda_{on}$ and $\lambda_{off}$ are calculated as
\begin{equation}
P_{OFF,OFF}(\Delta t) =  (1-\Lambda) + \Lambda  e^{-(\lambda_{on} + \lambda_{off})\Delta t} ,
\label{exp_off}
\end{equation} 
\begin{equation}
P_{ON,OFF}(\Delta t) = \Lambda  + (1-\Lambda) e^{-(\lambda_{on} + \lambda_{off})\Delta t} ,
\label{exp_on}
\end{equation} 
where $ \Lambda = \lambda_{off}/ (\lambda_{on}+\lambda_{off})$.

We calculate the probabilities $P_{OFF,OFF}$ and $P_{ON,OFF}$ for both HED and exponential distribution using  Eqs.~(\ref{eq:floatingequation}), (\ref{eq:floatingequation2}), (\ref{exp_off}) \& (\ref{exp_on}). For our analysis, we use HED and exponential distributions fitted for channel idle times of 2.4 GHz ISM band \cite{Stabellini2010}. The above probabilities for different PU load conditions are plotted in Figs.~\ref{fig:p11_04} and \ref{fig:p01_04}. As the inter-sensing interval $\Delta t$ increases, the conditional probabilities $P_{OFF,OFF}(\Delta t)$ and $P_{ON,OFF}(\Delta t)$ for exponential PU OFF times differs significantly. Further, we notice that these conditional probabilities converge to the stationary probabilities for an extremely long inter-sensing interval.

\section{Simplified structure of generalized predictive csa for channels with i.i.d. pu on-off times } \label{equivalence}
In this section, we will deduce the simplified structure of generalized predictive CSA framework in scenarios where  PUs on all channels have independent and identically distributed ON-OFF times. First, we prove the equivalence of Generalized Predictive CSA and POMDP-based Greedy CSA framework \cite{Zhao2007}, which were treated in CRN literature as a separate lines of work. Once their equivalence is proved, the interesting results of POMDP-based Greedy CSA will be applicable to Generalized Predictive CSA framework for channels with i.i.d. PU ON-OFF times. The following notations $p_{11,a}$ and $p_{01,a}$ represents the conditional probabilities $P_{OFF,OFF}^{a}$ and $P_{ON,OFF}^{a}$ of channel $a$,  respectively. 

\subsection{Generalized Predictive CSA: Slotted SU access}
We will briefly discuss the generalized predictive CSA framework where SU access the channel in slotted manner. By slotted SU access, we mean that if a secondary transmitter sense channel to be busy, it will again  sense new channel only at the beginning of next slot. 
At the beginning of each slot, the secondary user selects a channel which has the maximum probability of being free from PU. The actions taken by the secondary transmitter are as follows:

\begin{itemize}
\item Calculate the $\omega_{a}^{pr}(k)$ for every  channel $a$ at the beginning of the $k^{th}$ slot using the previous sensing results as
\begin{equation}
\omega_{a}^{pr}(k) = \begin{cases}
p_{11,a}(k-l), &\text{if }S_{a}(l)= 1.\\
p_{01,a}(k-l), &\text{if }S_{a}(l)= 0.  \nonumber
\end{cases} 
\end{equation}
where $p_{11,a}(n)$ and  $p_{01,a}(n)$ are the probabilities of channel $a$ being free from PU given the last sensing happens at $n$ slots ago. These probabilities can be calculated for a given PU ON-OFF time distributions using Eq.~\ref{eq:p11} \& 6. The symbol $l$ denotes the most recent slot in which channel $a$ is sensed and $S_{a}(l)$ denotes the sensing result of channel $a$ at the $l^{th}$ slot. $S_{s}(l)$ takes value one if the channel is idle and zero if it is busy. 

\item Select the channel $a_{*}$ that has the maximum probability of being idle at the beginning of the $k^{th}$ slot
\begin{equation}
a_{*}(k) = \arg\lbrace\max_{a=1,..,N_{c}} \omega_{a}^{pr}(k) \rbrace. \nonumber
\end{equation}

\item Sense the selected channel $a_{*}$. If channel is idle, the secondary user transmits frames in that slot. Otherwise, transmission is deferred in the slot

\end{itemize}

During the initial phase of Predictive CSA framework, the probability of channel being idle $\omega^{pr}_a(k)$ for channels which are not sensed once are assigned as
\begin{equation}
\omega_{a}^{pr}(k) = \omega_{a}^{pr}(0)p_{11,a}(k)  + (1-\omega_{a}^{pr}(0))p_{01,a}(k),
\label{eq:init_p}
\end{equation}
 where $\omega_{a}^{pr}(0)$ is the probability of channel $a$ being idle at the beginning of first slot.

\subsection{POMDP based Greedy CSA framework}

Greedy CSA for the multichannel CRN MAC protocol \cite{Zhao2007}  is described in four stages namely (i) Sensing Action (ii) Sensing Observation (iii) Access Action, and (iv) Updating Belief vector. 

\begin{itemize}
\item { Sensing Action: } Select the channel $a_{*}$ which has the maximum probability of being idle in the $k^{th}$ slot and sense the channel
\begin{equation}
a_{*}(k) = \arg\lbrace\max_{a=1,..,N_{c}} \omega_{a}^{gr}(k-1)p_{11,a} \\ + (1-\omega_{a}^{gr}(k-1)p_{01,a}) \rbrace \nonumber
\end{equation}
where $\omega_{a}^{gr}(k)$ be the probability of channel $a$ being free from PU at the beginning of the $k^{th}$ slot.

\item { Sensing Observation: }
Denote the sensing result of channel $a_{*}$ as $ S_{a_{*}}(k) = 1$ if the channel is free from PU. Otherwise let $ S_{a_{*}}(k)$ be zero.

\item { Access Action: }
 If $ S_{a_{*}}(k)=1 $, transmit the packet in the $k^{th}$ slot else defer transmission in that slot.

\item { Update Belief vector: }
\begin{equation}
\omega_{a}^{gr}(k) = \begin{cases}
\omega_{a}^{gr}(k-1)p_{11,a} + \\ (1-\omega_{a}^{gr}(k-1))p_{01,a}, &\text{if }a_{*}\neq a \\
				1, &\text{if } a_{*}=a, S_{a_{*}}(k)=1 \\
				0, &\text{if } a_{*}=a, S_{a_{*}}(k)=0. \label{eq:update}
			\end{cases}
\end{equation}

\end{itemize}

\subsection{Equality of Greedy and Predictive CSA frameworks}
The equality of Greedy and Predictive CSA framework will be proved using induction. Without losing generality, we define the initial belief vector $\Lambda(0) =~[\omega_{1}(0), \omega_{2}(0), ... \omega_{N_s}(0)]$ to be same for both frameworks. 

\newtheorem{t1}{Theorem}
\begin{t1}
The Predictive CSA framework and Greedy CSA framework are equivalent i.e. for $\forall n \geq 1$ , $\omega_{a}^{gr}(n) = \omega_{a}^{pr}(n)$. 
\end{t1}
\begin{proof} By mathematical induction, we can prove the above theorem for two cases namely (i) for channels which are not sensed even once, and (ii) for channels which are sensed at least once. \newline
\\ \textbf{ Case 1: } For a channel $a$ which is not sensed even once until $k^{th}$ slot. 

For $n=1$, $\omega_{a}^{gr}(1) = \omega_{a}^{pr}(1)$.
 (from Eq.~(\ref{eq:init_p}) \& (\ref{eq:update}) ) \\

Assume that $ \omega_{a}^{gr}(k) = \omega_{a}^{pr}(k)$ holds for $n=k$. \\

For $n=k+1$, 
the probability of a channel $a \neq a^{*}$ being free from PU at the beginning of the $(k+1)^{th}$ slot in Predictive CSA is given as
\begin{equation}
\begin{split}
\omega_{a}^{pr}(k+1) & = \omega_{a}(0)p_{11,a}(k+1) + (1-\omega_{a}(0))p_{01,a}(k+1)  
\\ & = \omega_{a}(0)[ p_{11,a}(k)p_{11,a} + p_{10,a}(k)p_{01,a}] +  
\\& \quad  (1-\omega_{a}(0))[p_{01,a}(k)p_{11,a} + p_{00,a}(k)p_{01,a}]
\\ & = \omega_{a}^{pr}(k)p_{11,a} + (1-\omega_{a}^{pr}(k))p_{01,a}
\\ & = \omega_{a}^{gr}(k)p_{11,a} + (1-\omega_{a}^{gr}(k))p_{01,a} = \omega_{a}^{gr}(k+1), \nonumber
\end{split}
\end{equation}
where the following facts are used
 \begin{equation}
 p_{11,a}(k+1) =  p_{11,a}(k) p_{11,a} + p_{10,a}(k)p_{01,a}, \nonumber
 \end{equation}
 \begin{equation}
 p_{01,a}(k+1) =  p_{01,a}(k) p_{11} + p_{00,a}(k)p_{01,a}. \nonumber 
 \end{equation}  
 
\textbf{Case 2 :} For a channel $a$ which is sensed at $l^{th}$ slot where $1 \leq l < k^{th} $ slot 

For $n=l+1$,
\begin{equation}
\omega_{a}^{gr}(l+1) = \begin{cases}
p_{11,a},  &\text{if }S_{a}(l)= 1\\ 
p_{01,a},  &\text{if }S_{a}(l)= 0. 
\end{cases}\nonumber
\end{equation}
\begin{equation}
\omega_{a}^{gr}(l+1) = \omega_{a}^{pr}(l+1). \nonumber
\end{equation}
Assuming $ \omega_{a}^{gr}(k) = \omega_{a}^{pr}(k)$ holds,
the probability of channel $a$ being free from PU at the beginning of the $(k+1)^{th}$ slot in  Greedy CSA  is given as
\begin{equation}
\begin{split}
\omega_{a}^{gr}(k+1) & = \omega_{a}^{gr}(k)p_{11,a} + [1-\omega_{a}^{gr}(k)]p_{01,a}
\\ & =\omega_{a}^{pr}(k)p_{11,a} + [1-\omega_{a}^{pr}(k)]p_{01,a}
%\\ & = \begin{cases}
%\text{if }S_{a}(l)= 1, \\ 
%p_{11,a}(k-l)p_{11,a} + (1-p_{11,a}(k-l))p_{01,a} \\ 
%\text{if }S_{a}(l)= 0 , \\
%p_{01,a}(k-l)p_{11,a} + (1-p_{01,a}(k-l))p_{01,a}
%\end{cases}
%\\ & = \begin{cases}
%\text{if }S_{a}(l)= 1,\\ 
%p_{11,a}(k-l)p_{11,a} + p_{10,a}(k-l)p_{01,a} \\
%\text{if }S_{a}(l)= 0, \\
%p_{01,a}(k-l)p_{11,a} + p_{00,a}(k-l)p_{01,a} 
%\end{cases}
\\ & = \begin{cases}
p_{11,a}(k+1-l)  ,&\text{if }S_{a}(l)= 1\\ 
p_{01,a}(k+1-l) ,  &\text{if }S_{a}(l)= 0 
\end{cases}
\\&  =  \omega_{a}^{pr}(k+1).
\end{split} \nonumber
\end{equation}

Hence by induction, it is proven that $\forall ~n ~\geq ~1 ~, ~\omega_{a}^{gr}(n) = \omega_{a}^{pr}(n)$. 
\end{proof}

\subsection{Simplified structure of Generalized Predictive CSA framework for i.i.d. PU ON-OFF times}

As we have proven the equivalence of predictive and greedy CSA framework, we will now use the properties of  greedy CSA framework to derive simplified structure of generalized predictive CSA framework for channels with i.i.d. PU ON-OFF times.

\begin{figure*}[ht]
\centering
  \includegraphics[scale=0.4, trim=1cm 0.75cm 0.75cm 1cm,clip=true]{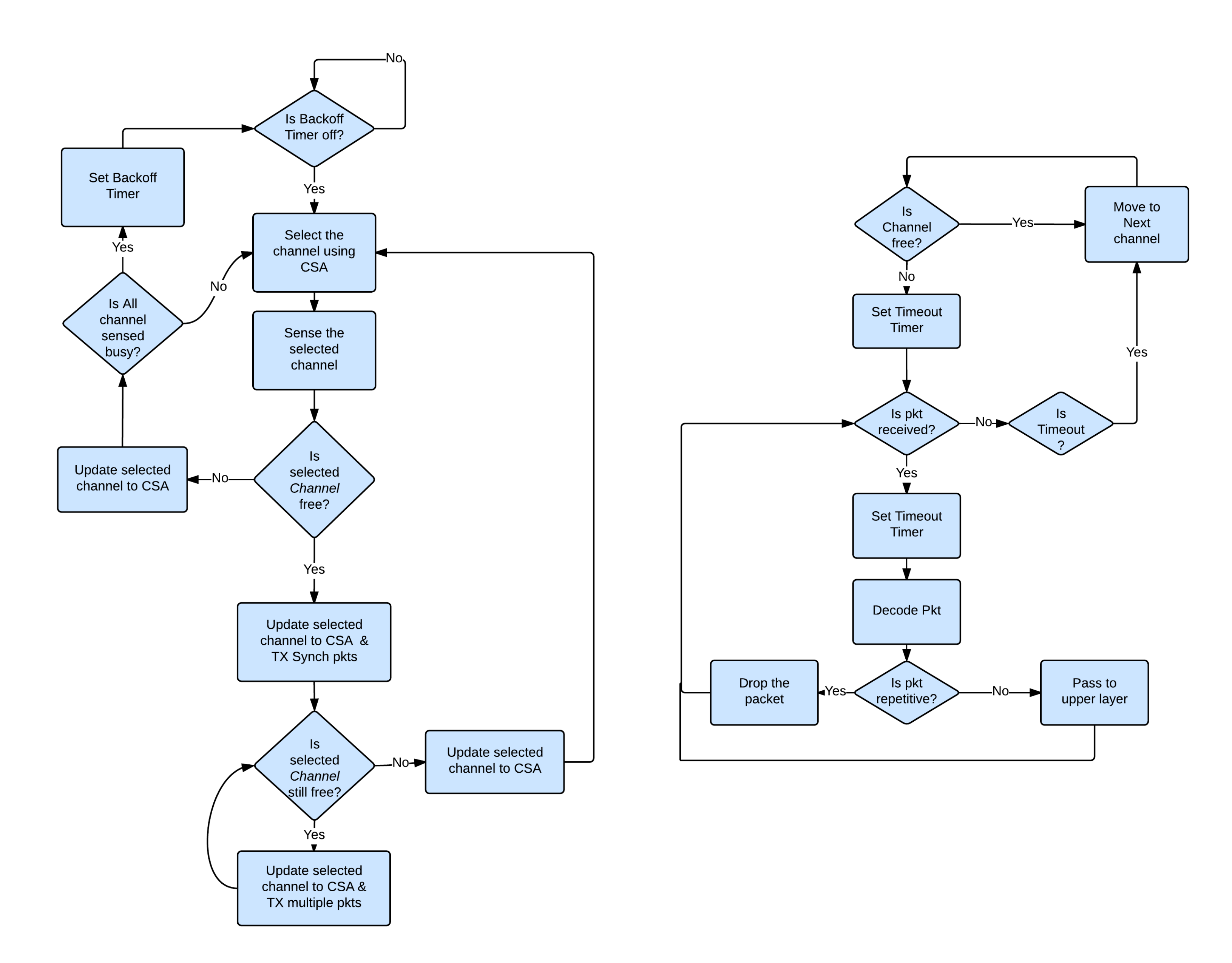} 
\caption{\small Medium Access Control (MAC) flowchart of secondary transmitter (left) and secondary receiver (right). }
\label{fig:flowchart}
\end{figure*}

\subsubsection*{Lemma 1}
When $N_{c}$ channels have i.i.d. PU ON-OFF times and the $P_{OFF,OFF}(\Delta t) \geq P_{ON,OFF}(\Delta t)$, the generalized predictive CSA is equivalent to round robin CSA where the ordering of channel is based on descending order of initial belief vector $\Lambda(0)$.
\vspace{-0.25cm}
\subsubsection*{Lemma 2}
When $N_{c}$ channels have i.i.d. PU ON-OFF times and the $P_{OFF,OFF}(\Delta t) < P_{ON,OFF}(\Delta t)$, the secondary transmitter, which uses generalized predictive CSA framework, switches to a channel most recently visited an even number of slots ago. 

\begin{proof}
The simplified structure of greedy CSA framework is proved in \cite{Zhao2008b} (Sec.IV, \textit{Structure of myopic sensing}) for $N_{c}$ i.i.d. channels. As we have proved the equality of greedy and generalized predictive CSA framework in Theorem 1, the Lemma 1 and Lemma 2 applies to the generalized predictive CSA framework. 
\end{proof}

Note that the exact values of $P_{OFF,OFF}(\Delta t)$ and $P_{ON,OFF}(\Delta t)$ are not required in above scenarios as long as their ordering is preserved, i.e.~( $P_{OFF,OFF}(\Delta t) < P_{ON,OFF}(\Delta t)$ or $P_{OFF,OFF}(\Delta t) \geq P_{ON,OFF}(\Delta t)$ ). However, the generalized predictive CSA framework plays a vital role when channels have non-identical idle time distributions.  

\subsection{Computational Complexity of CSAs}
The  proposed   Generalized   Predictive CSA  and   Predictive   CSA   uses   the   same   CSA   framework.  
The   main   difference   between   them   is   only   in   the   calculation   of   probability   of   channel   being  idle   given   the   previous   sensing   result,   i.e.   $P_{OFF,OFF} (\Delta t)$ and  $P _{ON,OFF}(\Delta t)$, used   by   CSA  algorithms. The   Generalized   Predictive   CSA   for   HED   OFF   times   uses   Eq.~(11) and~(12)  whereas a Predictive   CSA   uses   Eq.~(13) and~(14).   In   the   calculation   of   probabilities   $P^a_{OFF,OFF}(\Delta t_a )$ ,  $P^a_{ON,OFF}(\Delta t_a)$ $\forall$ $a  \in [ 1, N_c ]$  given   in   Eq.~(11)--Eq.~(14),   only   the   exponential   term   is   a  function   of   inter-sensing   duration  $\Delta t_a$.   Hence,   the   computational   complexity   of   proposed  Generalized Predictive CSA is same as that of Predictive CSA.  In terms of Big­O notation, both CSA algorithms have computational complexity of  $\mathcal{O}(N_c)$.

\section{Simple Multi-channel CRN MAC}
\label{sec:MAC_design}
In this section, we will design a simple multichannel CRN MAC protocol to evaluate the performance of the proposed channel selection algorithm. We consider single secondary transmitter-receiver pair without assuming any dedicated control channel between the transmitter and the receiver. We have adopted a rendezvous mechanism given in \cite{Ansari2013} for our work to easily synchronize secondary transmitter and receiver on a new channel whenever PU reappears on a current operating channel.

\subsection{Transmitter}
We assume saturated secondary transmitter that always has frames in the transmit buffer.  Let $t_{frame}$ be the frame transmission time, $t_{inter}$ be the time interval between two subsequent frames, and $t_{sense}$ be the channel sensing time. Since we assume half-duplex secondary system, as is the case for existing transceivers, the secondary transmitter has to switch from TX mode to RX mode for sensing the channel. Similarly, it has to switch back to TX mode for frame transmission. Let the switching times be denoted by $t_{RXmode}$ and $t_{TXmode}$, respectively. The secondary transmitter sense the channel for $t_{sense}$ duration in RX mode. If a channel is idle, the secondary transmitter switch to the TX mode and starts transmitting the frames for $t_{PUallow}$ duration. The number of frames, $N_{frame}$, that can be transmitted in $t_{PUallow}$ is calculated as 
\begin{equation}
 N_{frame}=~ \lfloor\frac{(t_{PUallow} - t_{RXmode} -t_{TXmode})}{(t_{frame}+t_{inter})} \rfloor. \nonumber
 \end{equation} 
 
After the transmission of $N_{frame}$ frames, the secondary transmitter once again switches back to the RX mode for sensing. When the channel is sensed busy, the secondary transmitter immediately shifts to a new channel, provided by  CSA, for sensing. If all the channels are sensed busy, the transmitter goes to the \emph{Backoff} state and remains idle for fixed $t_{Backoff}$ time. The \emph{Backoff} mechanism avoids unnecessary channel switching as the probability of channel moving from busy to idle state in a short duration is very small. 
\subsection{Secondary transmitter-receiver synchronization}  \label{sec:rendezvous}
When secondary transmitter shifts to a new channel, the secondary receiver has to be informed either explicitly through a dedicated control channel or implicitly by a \emph{rendezvous algorithm} \cite{Romaszko2013}. In a more realistic cognitive network scenarios, it is often difficult to ensure a common control channel and time synchronization between transmitter and receiver. For our work, we use a rendezvous algorithm for transmitter-receiver channel synchronization. We compare the performance of our proposed CSA with predictive CSA in terms of \textit{channel switch rate} at the secondary transmitter, which is independent of the choice of rendezvous algorithm. For ease of implementation, we have chosen \emph{CogMAC} rendezvous algorithm of Ansari \textit{et al.} \cite{Ansari2013}, which we briefly explain in the following section. However, this distributed algorithm trades off part of the bit rate efficiency to ensure reliability of the rendezvous. We refer readers to \cite{Romaszko2012} for a review on asynchronous rendezvous algorithms.

\subsection{Receiver} \label{sec:receiver}
The secondary receiver senses each channel consecutively for $t_{sense}$ duration. When there is no transmission from either the PU or secondary transmitter in a channel, the secondary receiver senses channel to be idle and moves to a new channel. When a channel is sensed busy, the secondary receiver locks to that channel and demodulates the received frames to identify the transmission source (PU or secondary transmitter). When the received frames are not decodable, secondary receiver assumes channel to be occupied by PU and shifts to other channel. If the decoded frames are addressed to the secondary receiver, they are placed in the MAC receive buffer. Secondary receiver's MAC fires \emph{Timeout} timer and polls receive buffer whenever the channel is sensed busy. If receive buffer has any frames within $t_{timeout}$ after locking to  a particular channel, it stays in that channel and resets the \emph{Timeout} timer as shown in Fig.~\ref{fig:flowchart}. Suppose the MAC receive buffer does not have any frame during the $t_{timeout}$ duration, then secondary receiver assumes that the secondary transmitter has moved on to a new channel and starts searching other channels for secondary frames. This assumption may be wrong as the secondary frames may be lost due to noise or primary user interference. Though these errors cannot be completely avoided in wireless environment, we can reduce those by choosing proper $t_{timeout}$ duration:
\begin{equation}
t_{Timeout}\geq \max \lbrace 3t_{frame},t_{TXmode}+t_{RXmode}+t_{sense}+t_{frame} \rbrace. \nonumber
\end{equation}
To minimize the initial frame loss during the rendezvous process, i.e. meeting of secondary transmitter and receiver on a same vacant channel, the secondary transmitter repeats the initial frame whenever it shifts to a new channel. The optimal number of initial frame repetition, which is a function of MAC layer timing parameters, and other details of our \textit{rendezvous} algorithm are given in our previous works \cite{Ansari2010}, \cite{Ansari2013}. The MAC level flowchart of secondary transmitter and secondary receiver is shown in Fig.~\ref{fig:flowchart}.

\subsection{Synchronization overhead }
In the \textit{CogMAC rendezvous} algorithm, the overhead time in transmitter--receiver synchronization is calculated based on the time duration $t_{repeat-time}$, during which the initial frame has to be repeated to avoid packet loss in secondary network’s channel switching. Therefore, the initial frame repetition time in transmitter--receiver synchronization should be atleast $t_{repeat-time} \geq N_c t_{sense} +(N_c -1)t_{switch}$. For example, in our test-bed experiment with  $N_c=4$, $t_{sense}= 40ms$, $t_{switch}=25ms$, we get $t_{repeat-time} =235 ms$. Since the $t_{frame} = 200ms$, the initial frame has to be repeated atleast twice to avoid packet loss at secondary receiver during channel switch at transmitter side. This results in minimum overhead of $2*t_{frame}= 400ms$ in transmitter--receiver synchronization in our test-bed implementation.

\section{Performance evaluation of CSA} \label{experiment}

The Generalized Predictive CSA framework is incorporated on a secondary transmitter in multichannel CRN MAC protocol. We experimentally evaluate the performance of the proposed CSA on a Software Defined Radio (SDR) based wireless test-bed. The wireless test-bed consists of multiple Ettus Research USRP2 devices \cite{Ettus} connected to PCs, which are installed with NI LabVIEW \cite{Labview}. The Physical (PHY) layer algorithms such as frame detection and synchronization developed for CogMAC+ protocol \cite{Wang2014} are reused in our test-bed. However, we have implemented a simple multichannel CRN MAC protocol detailed in Section~\ref{sec:MAC_design} on the test-bed. The channel selection algorithm and MAC protocol for secondary network are programmed in NI LabVIEW using Decomposable MAC framework \cite{Ansari2010}. 

\begin{figure}[ht]
\centering
 \includegraphics[scale=0.48, trim=4mm 5mm 0mm 0mm,clip=true]{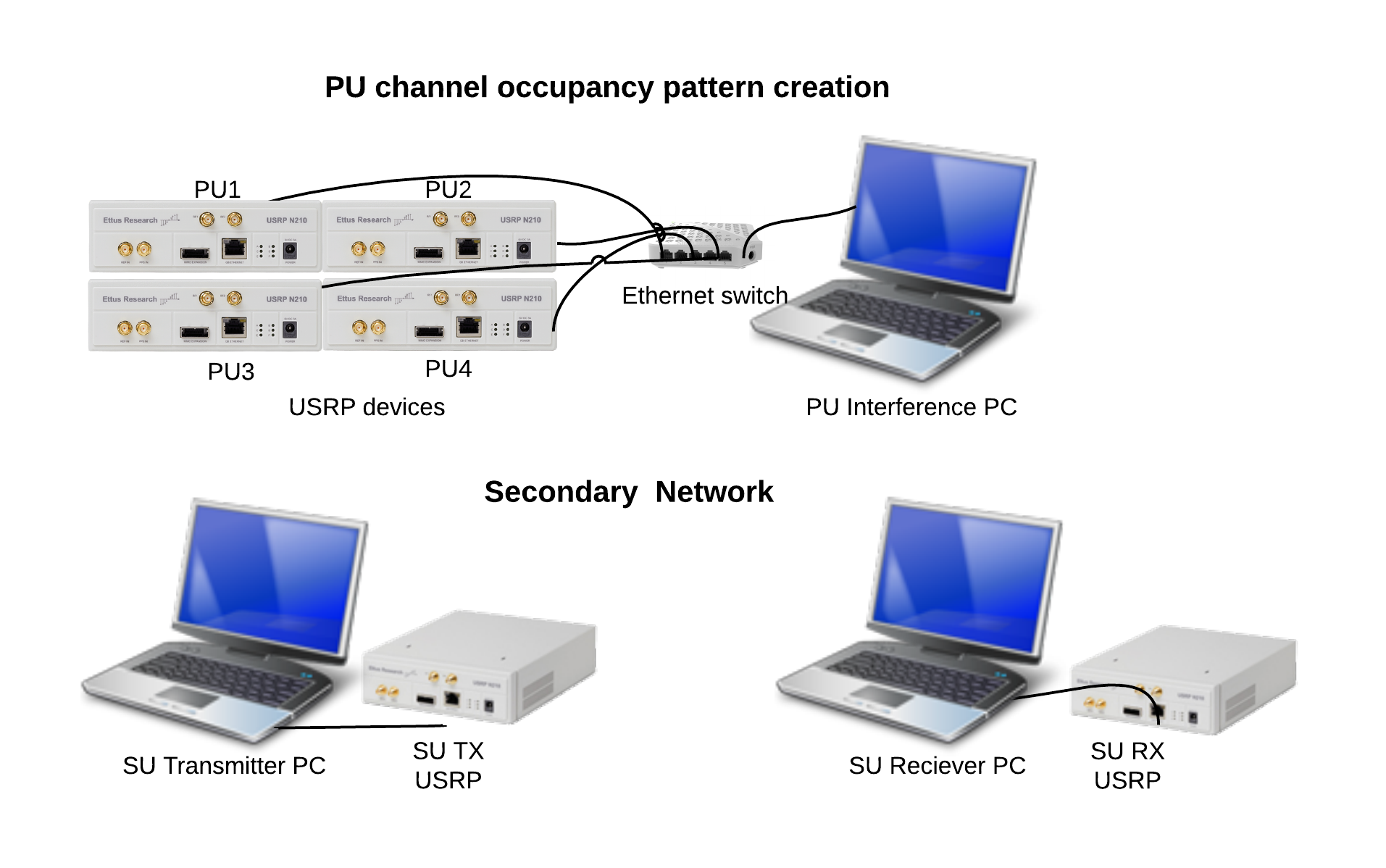} 
\caption{Schematics of wireless test-bed used in the performance evaluation of proposed CSA in a realistic environment. }
\label{fig:testbed}
\end{figure}
\begin{figure*}[ht]
\begin{minipage}[b]{0.3\linewidth}
\centering
\includegraphics[scale=0.43, trim=7mm 0mm 7mm 7mm,clip=true]
{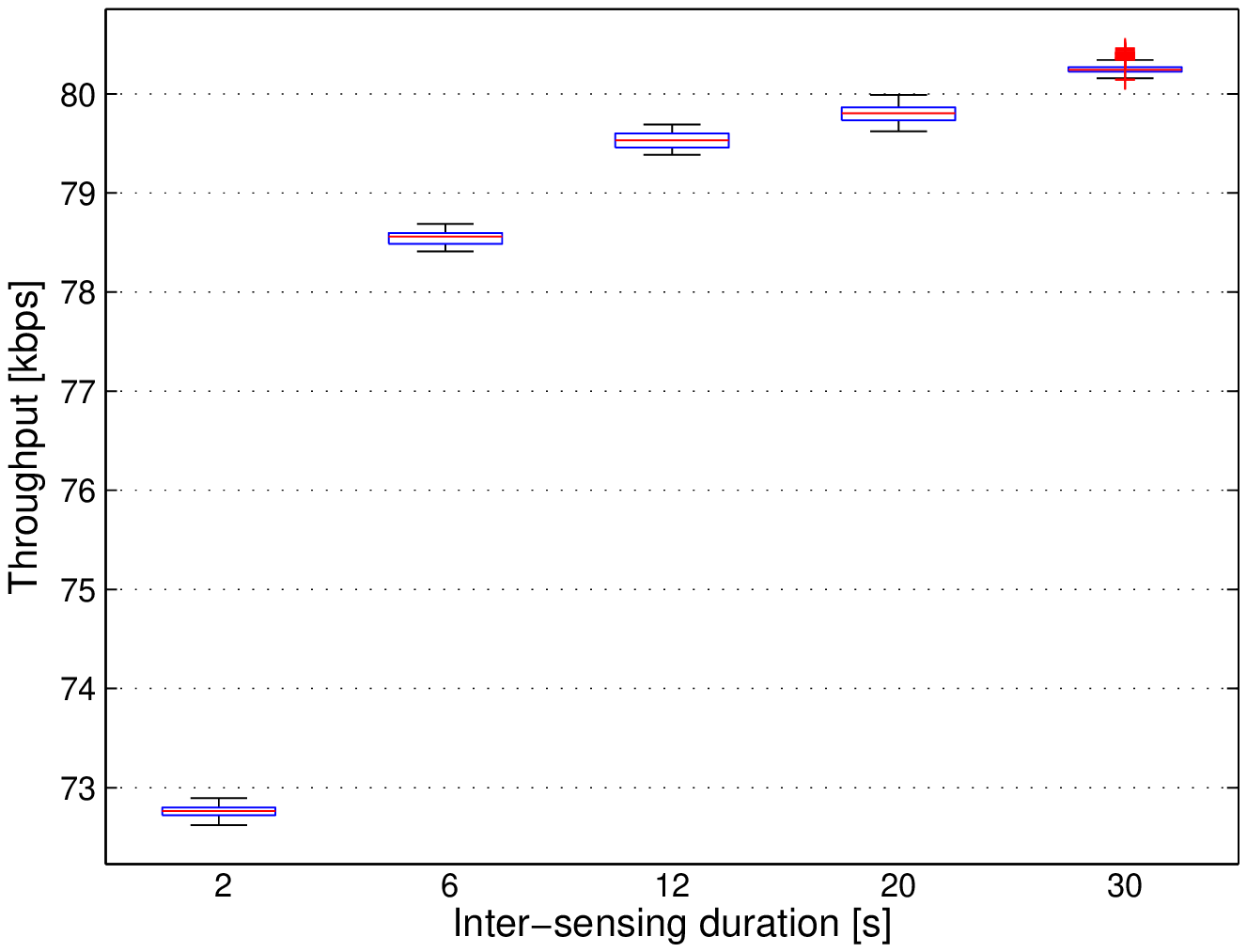}
\caption{\small The box plot of secondary receiver's throughput for different inter-sensing duration without any PU activity on a channel- Benchmark reference of wireless test-bed.}
\label{fig:base_diff_inter_sensing}
\end{minipage} 
\hfill
\begin{minipage}[b]{0.3\linewidth}
\centering
\includegraphics[scale=0.43, trim=7mm 0mm 7mm 7mm,clip=true]
{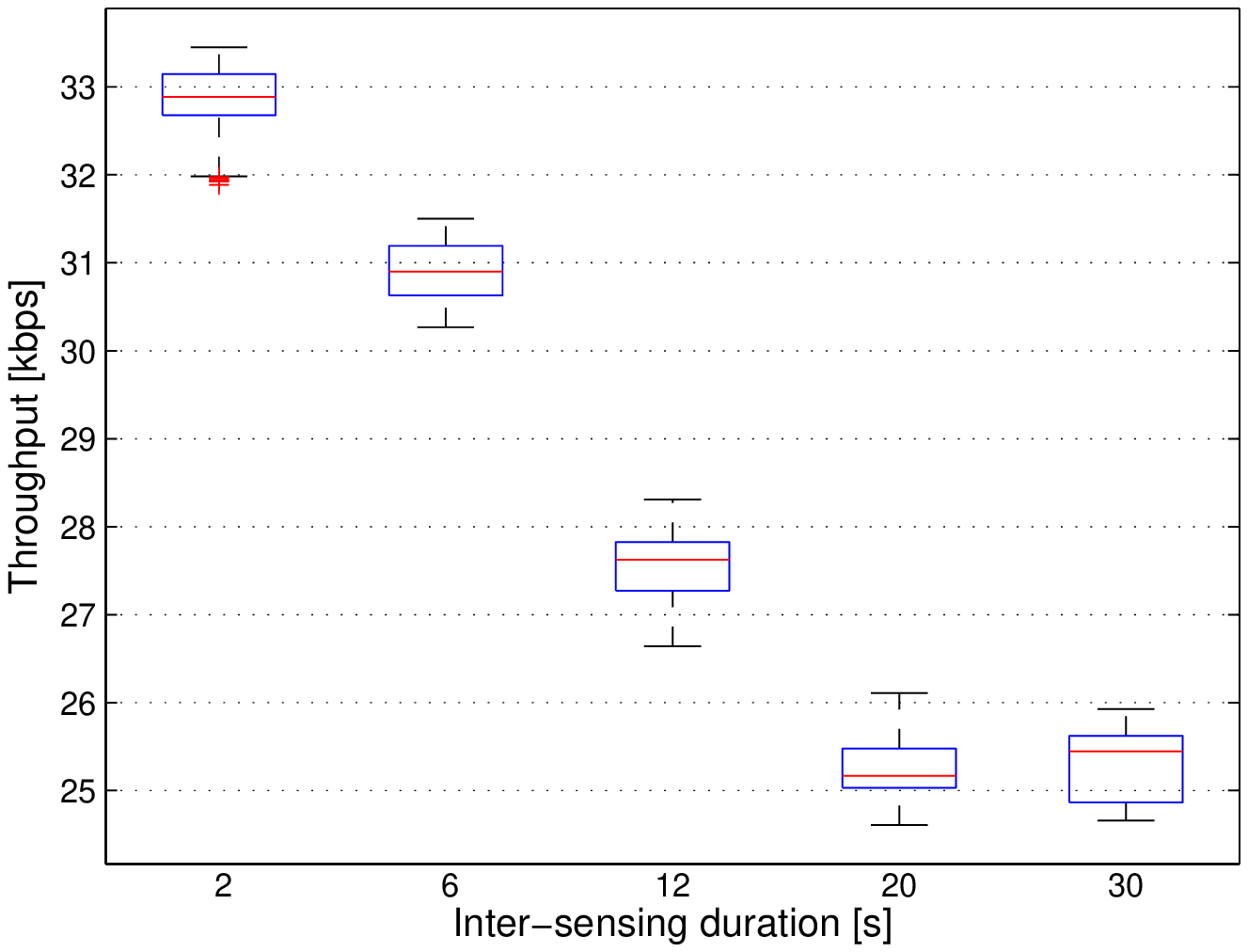}
\caption{\small The box plot of secondary receiver's throughput for different inter-sensing duration with the same PU duty cycle of 0.5 and HED OFF times on all four channels. }% with $t_{sense} = 100 ms$ }
\label{fig:interf_sys_05_inter_sen}
\end{minipage}
\hfill
\begin{minipage}[b]{0.3\linewidth}
\centering
\includegraphics[scale=0.43, trim=7mm 0mm 0mm 7mm,clip=true]
{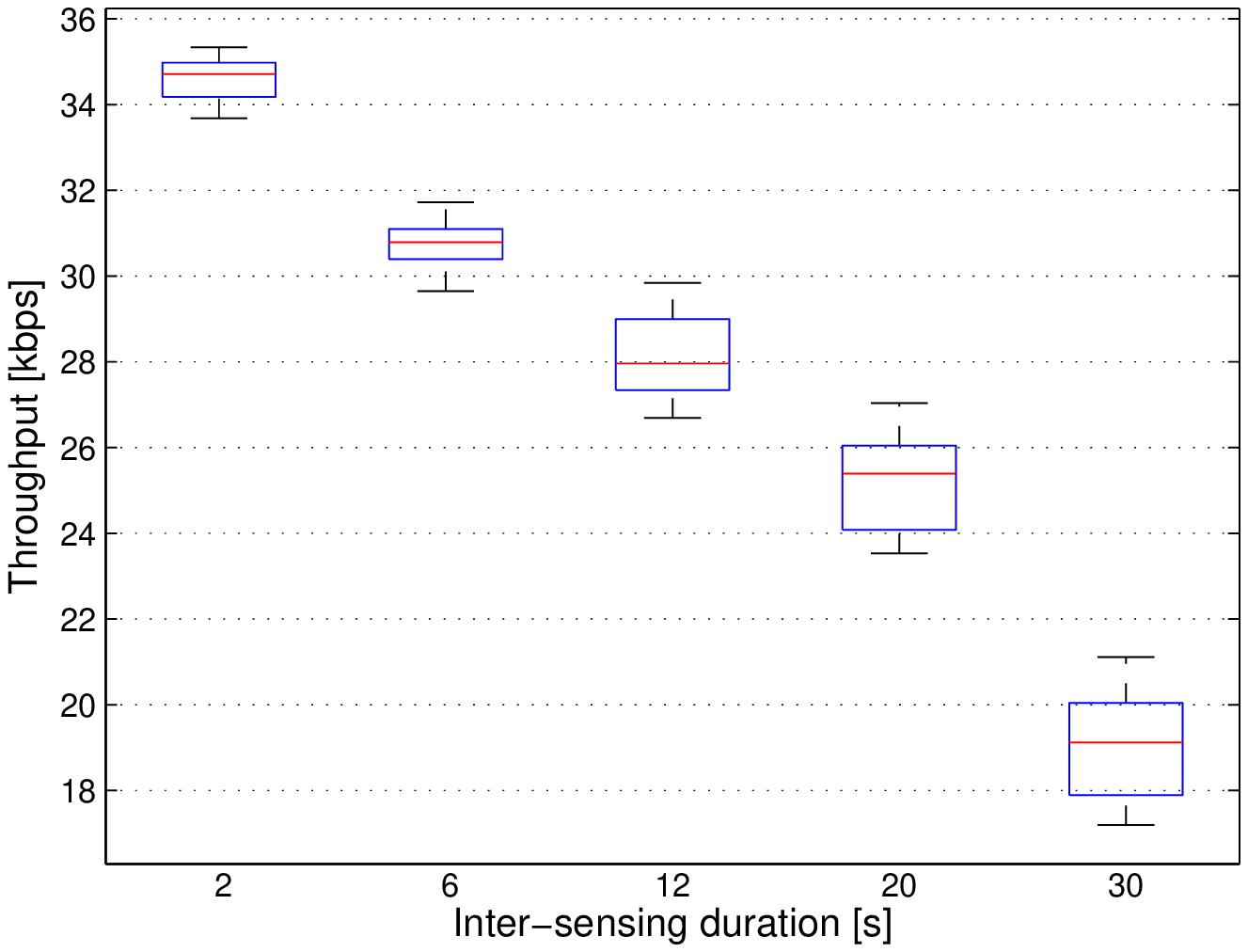}
\caption{\small The box plot of secondary receiver's throughput for different inter-sensing duration with PU duty cycle of [0.7 0.2 0.4 0.7] and HED OFF times on all four channels.}
\label{fig:interf_sys_05_inter_sens}
\end{minipage}

\end{figure*}

\subsection{Test-bed to evaluate proposed CSA}
The experimental wireless test-bed consists of separate secondary transmitter and receiver PCs, each connected to a USRP2 device with a radio front end which can operate from 400 MHz to 4.4 GHz. The PHY and MAC parameters of the wireless test-bed are listed in Table ~\ref{table:params}. The values of MAC layer parameters such as $t_{RXmode}$, $t_{TXmode}$ and $t_{switch}$ are assigned based on the hardware profiling of USRP devices \cite{Online} and the processing delays in NI LabVIEW.
\begin{table}[ht]
\caption{Values of PHY \& MAC parameters used in our experiment. We have used QPSK modulation scheme.}
\centering
\tiny
\begin{tabular}{c c |c|c|c}
\hline\hline  
 & parameters & value & parameters & value\\[1ex] 
\hline
& $t_{TXmode}$ & 15 ms & $t_{RXmode}$ & 150 ms \\
&$t_{frame}$ &  200 ms &$t_{Timeout}$ & 700 ms \\
&$t_{switch}$ & 25 ms  &$t_{Backoff}$ & 4 ms\\ 
&Bandwidth & 400 kHz &  &  \\ [1ex]
\hline
\end{tabular}
\label{table:params}
\end{table}

The PU interference is generated using an independent LabVIEW application, which runs on a separate PC labelled as \emph{PU Interference PC} as shown in Fig.~\ref{fig:testbed}. In our experiment, we evaluate the performance of CSA by recreating PU's channel occupancy pattern fitted for realistic measurement traces \cite{Stabellini2010}. We use the parameters of ON-OFF time distributions (exponential and HED) given in \cite{Stabellini2010} for generating PU interference. However, we scaled up the distribution parameters by a factor of ten thousand  in our experiment for the following reasons:
\begin{itemize}
\item To make system stable, the experimental inter-sensing duration has to be on the order of  seconds because the frame time, receiver timeout, and other switching parameters of USRP system are on the order of several milliseconds.
\item To avoid the impact of processing delays of PHY and MAC layer algorithms in NI LabVIEW
\item To avoid the influence of other parameters like 'Backoff' at the transmitter on the performance differences of CSAs.
\end{itemize}

In our experiment, we have created PU channel occupancy patterns on four channels, with four USRP2 devices connected to PC through Gigabit Ethernet switch as shown in Fig.~\ref{fig:testbed}. We will now define the performance metrics such as average throughput and channel switch rate which are used to evaluate the performance of proposed CSA with existing CSA frameworks.
\subsubsection*{Throughput}
 The throughput of secondary network $\Gamma(t)$ is defined as the ratio of product of frame size (in bits) and the number of successfully received frames till time $t$ to the elapsed experimental duration $t$.
 \begin{equation}
 \Gamma(t) = \frac{N_F(t) * \textsf{Frame size [bits]} } {t}, \nonumber
 \end{equation}
where $N_F(t)$ denotes the number of correctly decoded frames at the secondary receiver till time $t$. The average throughput of secondary network $\Gamma$ is given by
 \begin{equation}
 \Gamma = \lim_{t \to \infty} \Gamma(t) \nonumber
 \end{equation}

\subsubsection*{Channel Switch Rate}
The channel switch rate, $C(t)$, is defined as the ratio of number of channel switches at the secondary transmitter to the time period $t$. The reason for choosing 'channel switch rate' as performance metric is to avoid the impact of rendezvous algorithm at the receiver on the performance study of CSA. 

\subsection{Discussion of experimental results}
We evaluate the throughput of wireless test-bed without PU activity on a channel and this acts as a benchmark reference for our forthcoming experiments. Although channel sensing is not required in the above scenario, we introduce periodic channel sensing in the benchmark reference to account for TX/RX mode switch delays and other processing delay in our test-bed. The average throughput of wireless test-bed (our benchmark reference) without PU activity on a channel is plotted against different inter-sensing duration ($\simeq t_{PUallow}$) in Fig.~\ref{fig:base_diff_inter_sensing}. We have used boxplot function in MATLAB to plot the secondary receiver's throughput. The red line in the box plot represents the median throughput. The upper and lower edge of blue box represents the 25th and 75th percentiles, respectively. In the absence of PU interference, the average throughput of secondary network degrades with decrease in inter-sensing duration as expected.

We will now evaluate the performance of generalized predictive CSA framework in the case of PU occupying channels intermittently. We consider four primary channels with exponential ON and HED/exponential OFF time distributions. 
We calculate the average throughput of secondary network in the following two scenarios: In the first scenario, PUs have 0.5 duty cycle and HED OFF times on all four channels. In the second scenario, we consider four channels having different PU duty cycle and with HED OFF times. The average throughput of the secondary network for both the scenarios are plotted in Figs.~\ref{fig:interf_sys_05_inter_sen} and~\ref{fig:interf_sys_05_inter_sens}.
On contrary to interference-free channels, we have observed that throughput of secondary network decreases with increase in inter-sensing duration. As the inter-sensing duration increases, the probability of collisions between secondary and primary frames also increases which results in lower throughput for secondary network.

\begin{figure}[ht]
\centering
\includegraphics[scale=0.5,trim=0mm 0mm 0mm 0mm,clip=false]{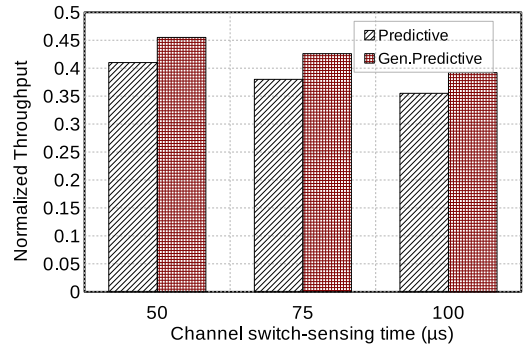}
\caption{\small Normalized throughput of Generalized Predictive and Predictive CSAs with PU duty cycle of [0.9 0.7 0.9 0.6 0.8 0.9] and HED OFF times on all six channels. The inter-sensing duration is set to 0.3 ms}
\label{fig:thg_comparison}
\end{figure}

In our experimental test-bed, the channel switching and sensing time together is on the order of $40 ms$. As we have scaled-up the PU ON-OFF times to operate test-bed in stable condition, the average PU OFF time and inter-sensing duration $t_{PUallow}$ are in the order of seconds. Hence, we do not find remarkable differences in the throughput of Generalized Predictive and Predictive CSA frameworks from the experimental traces. We note also that the throughput gain of proposed CSA might not be very pronounced because of a rather simple \textit{rendezvous} algorithm used at secondary receiver in our CRN MAC protocol. 

Hence, we have carried out MATLAB based simulation without scaling up HED parameters  \cite{Stabellini2010} of PU OFF times.  We have evaluated the normalized throughput of Generalized Predictive and Predictive CSA framework with PU duty cycle of [0.9 0.7 0.9 0.6 0.8 0.9] and HED OFF times on all six channels. We assume that the secondary receiver has perfect knowledge on the operating channel of secondary transmitter and hence avoid the impact of rendezvous algorithm on throughput.  We have plotted the normalized throughput of Generalized Predictive and Predictive CSA frameworks for various channel switch-sensing times in Fig.~\ref{fig:thg_comparison}. It can be noticed that the Generalized Predictive scheme has about 12\% higher throughput than the Predictive CSA.

\subsubsection*{Channel switch rate}
We will now evaluate the performance of Generalized Predictive and Predictive CSA frameworks in terms of channel switch rate at the secondary transmitter. The \textsl{'Channel switch rate'} is chosen as a performance metric for benchmarking CSAs because  of its robustness to errors in wireless environment. For example, the average secondary network throughput metric is not robust against fading and transmission errors in wireless channel. Further, the average throughput  of the secondary network depends on the efficiency of rendezvous algorithm at the secondary receivers. 

\begin{figure}[ht]
 \centering
 \subfloat[Inter-sensing duration of 1000 ms]
 {{
\includegraphics[scale=0.45,trim=6mm 0mm 0mm 0mm,clip=true]{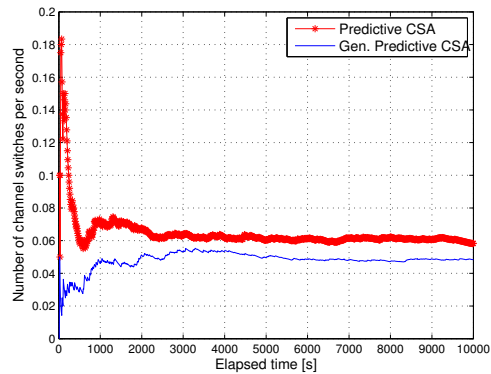}
\label{fig:perf_gain_1s}
}}
\hfill
\subfloat[Inter-sensing duration of 3000 ms]
{{
\includegraphics[scale=0.45,trim=6mm 0mm 0mm 0mm,clip=true]{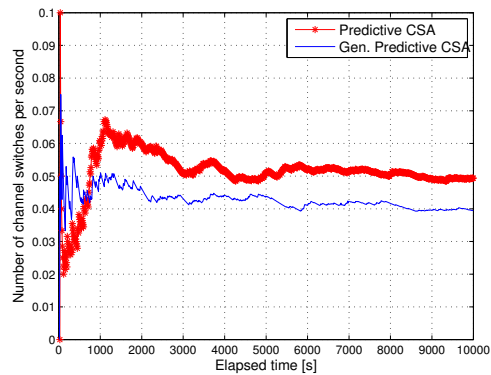}

\label{fig:perf_gain_3s}
}}
\caption{\small  Channel switching rate of Generalized Predictive CSA and Predictive CSA with PU duty cycle of 0.3 on four channels for inter-sensing duration of 1000 ms and 3000 ms.}
\label{fig:perf_gain}
\end{figure}

For evaluation on  experimental wireless test-bed, we consider a scenario where there are four primary channels with the same PU duty cycle value of 0.3. However, PUs on channels 1 and 4 have HED OFF time distribution with scaled-up parameters whereas PUs on channels 2 and 3 follow exponential OFF time distribution. The \textsl{Channel switch rate} at the secondary transmitter for Generalized Predictive CSA and Predictive CSA frameworks are plotted in Fig.~\ref{fig:perf_gain} for inter-sensing duration of 1000 ms and 3000 ms. The initial peaks  observed in the plots corresponds to the learning phase involved in Adaptive sensing threshold algorithm implemented in PHY layer.  Figs.~\ref{fig:perf_gain} (a) and (b) clearly show that the channel switch rate of Generalized Predictive CSA is significantly lower than the Predictive CSA. The Generalized Predictive CSA framework reduces channel switch rate by 16.7\% and 20\%  as compared to Predictive CSA for inter-sensing duration of 1000 ms and 3000 ms, respectively.
\begin{figure}[ht]
\centering
\includegraphics[scale=0.45,trim=0mm 0mm 0mm 0mm,clip=true]{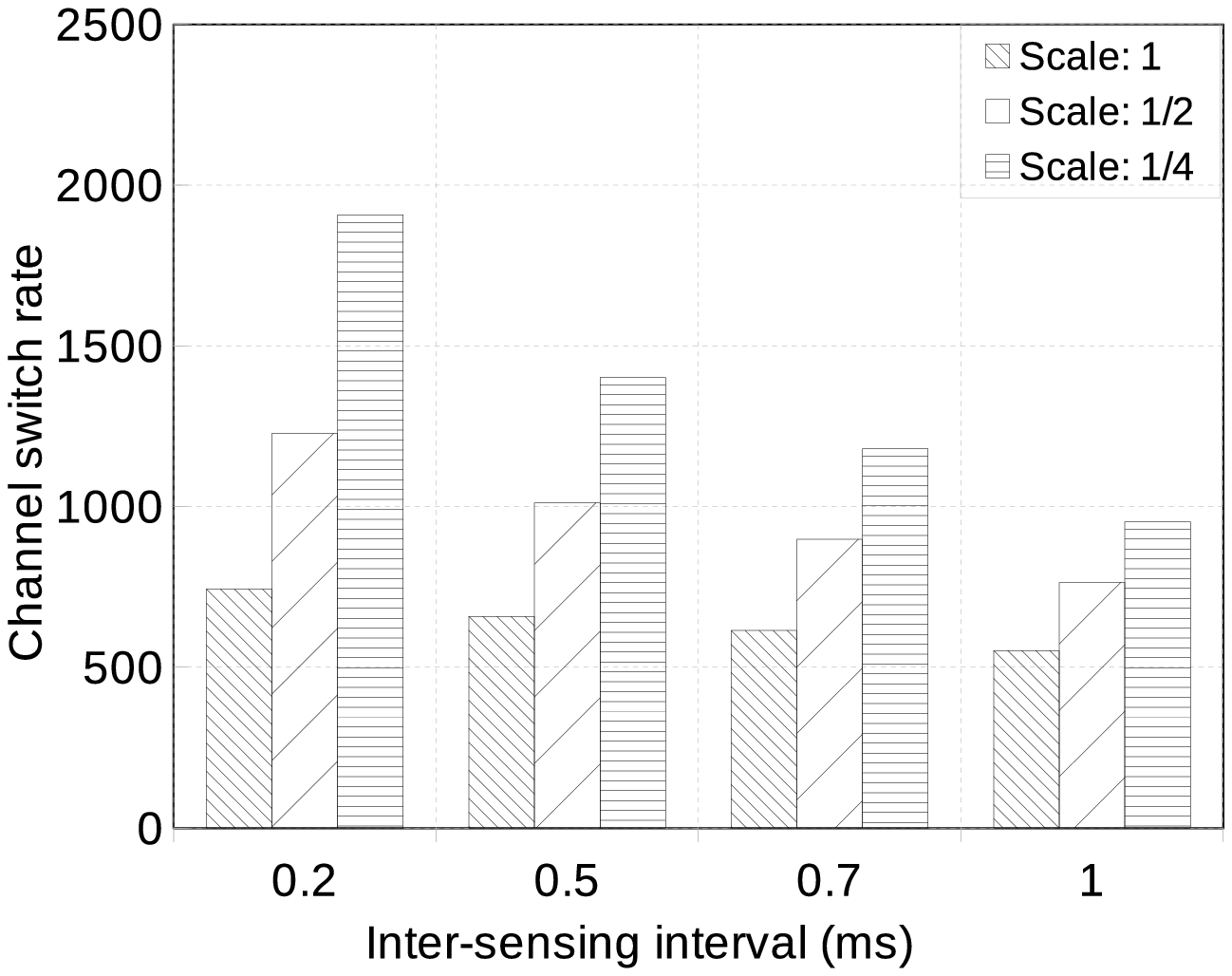}
\caption{\small Channel switch rate of Generalized Predictive CSA with 0.8 PU duty cycle on all six channels with HED OFF times on channels 1, 2, 4, 5 and exponential OFF times on channels 3, 6 for varying scale of PU duty cycle length.}
\label{fig:csr_comp}
\end{figure}
\begin{figure}[ht]
\centering
\includegraphics[scale=0.45,trim=0mm 0mm 0mm 0mm,clip=true]{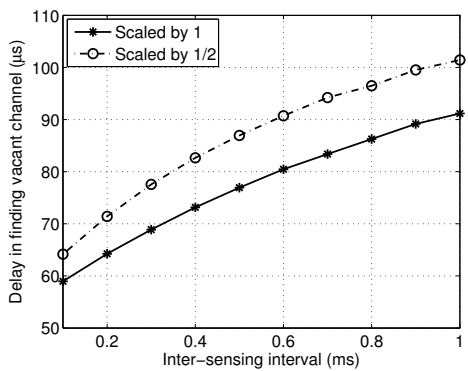}
\caption{\small The delay in finding the first vacant channel with 0.8 PU duty cycle on all six channels with HED OFF times on channels 1, 2, 4, 5 and exponential OFF times on channels 3, 6. The HED parameters of OFF times are taken from \cite{Stabellini2010} with length of PU duty cycle scaled by a factor of $1$ and $1/2$. The channel sensing time $t_{sense}$ is set to 50$\mu$s.}
\label{fig:delay}
\end{figure}
 
\begin{figure*}[ht]
\centering
\subfloat[For HED parameters in \cite{Stabellini2010} ]
{{
\includegraphics[scale = 0.45, trim=0mm 0mm 2mm 0mm,clip=true]{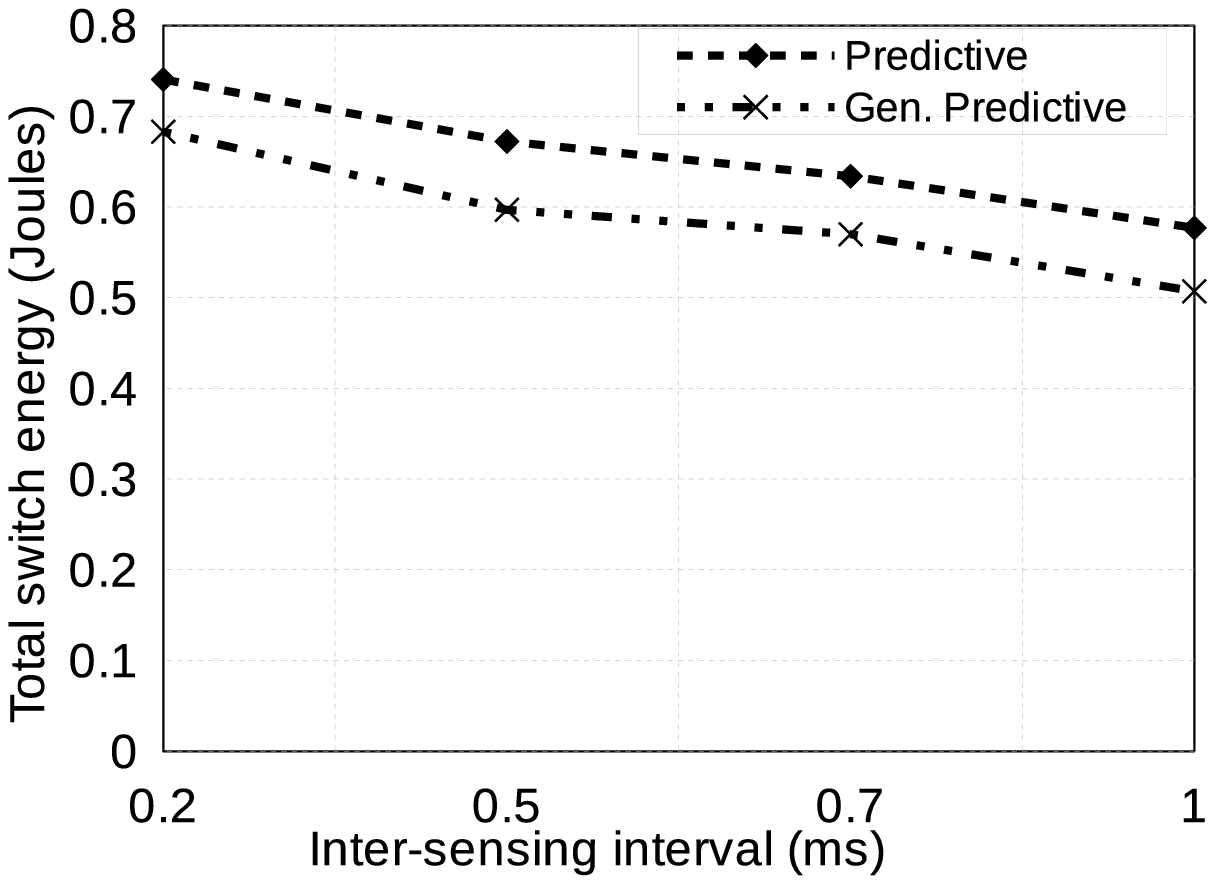}
}}
\hfill
\subfloat[For HED parameters in \cite{Stabellini2010} with length of PU duty cycle scaled by $1/2$]
{{
\includegraphics[scale = 0.45, trim=0mm 0mm 2mm 0mm,clip=true]{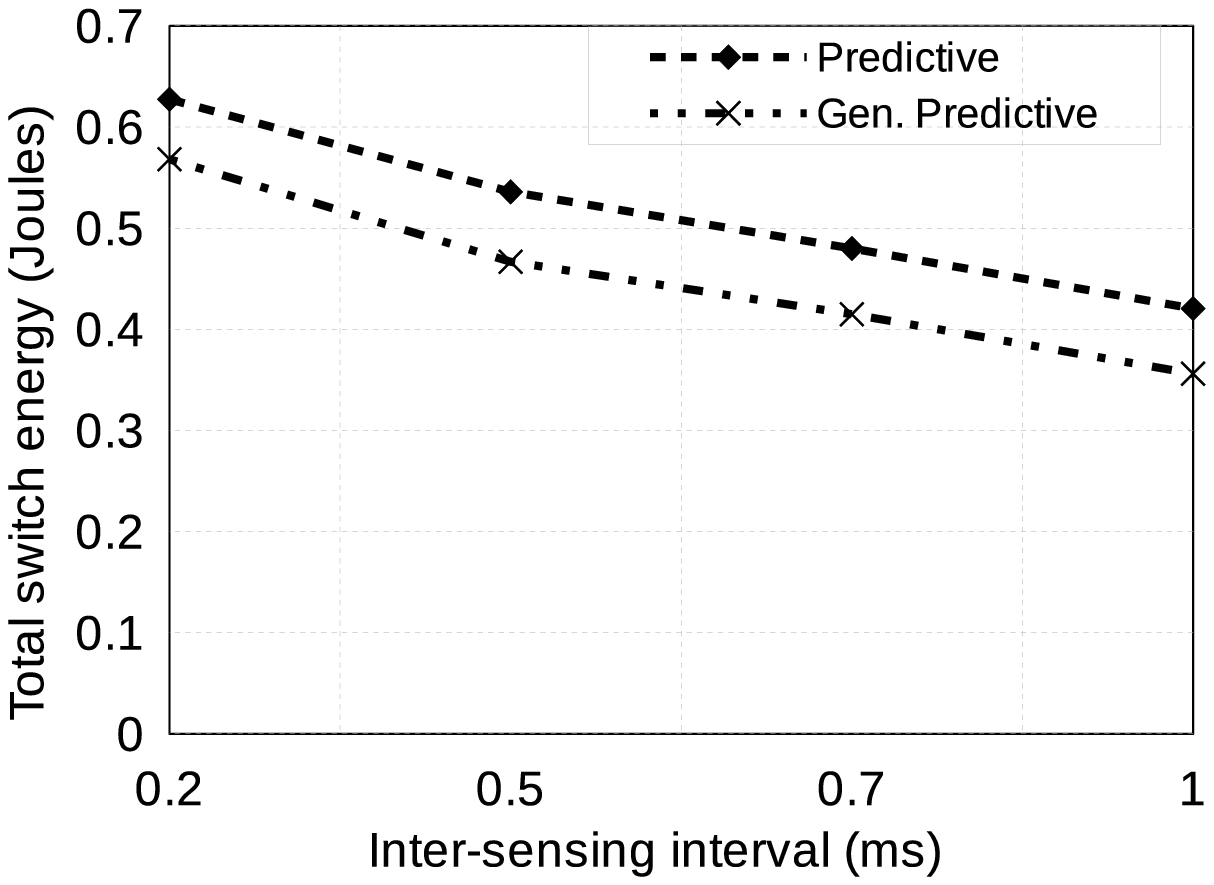}
}}
\hfill
\subfloat[For HED parameters in \cite{Stabellini2010} with length of PU duty cycle scaled by $1/4$]
{{
\includegraphics[scale = 0.45, trim=0mm 0mm 2mm 0mm,clip=true]{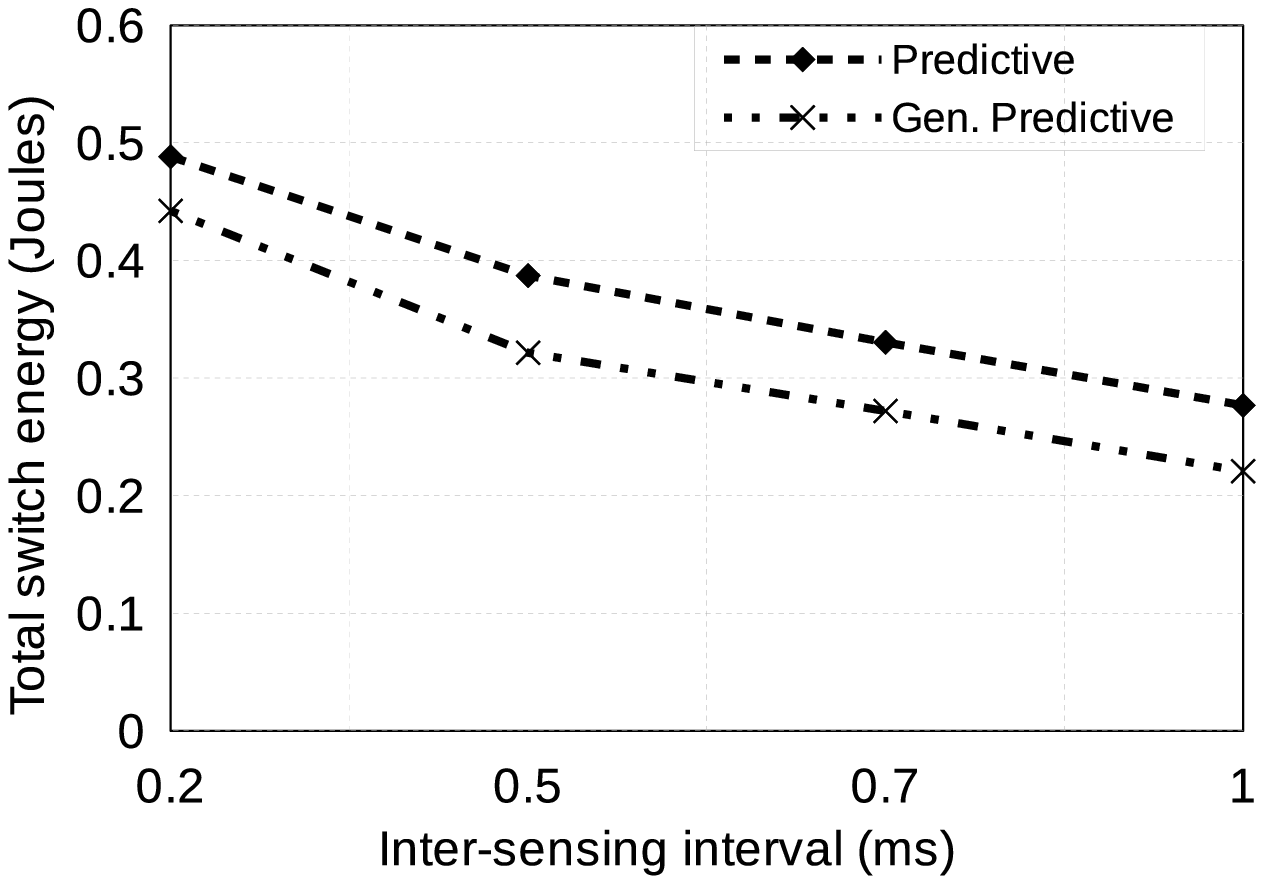}
}}
\caption{\small Total channel switch energy of Generalized Predictive and Predictive CSAs with 0.8 PU duty cycle on all six channels with HED OFF times on channels 1, 2, 4, 5 and exponential OFF times on channels 3, 6. The energy consumption is calculated for 10,000 PU ON-OFF cycles.} 
\label{fig:plot1}
\end{figure*}

Note that the PU ON-OFF time distribution parameters are scaled up by a factor of 10,000 for stability of experimental wireless test-bed. Hence, if we scale down parameter and inter-sensing duration to realistic values accordingly, the difference in number of channel switches for Generalized Predictive CSA and Predictive CSA framework are  significant. The MATLAB simulation is carried out to plot the channel switch rate of Generalized Predictive CSA in realistic conditions without scaling up distribution parameters by a factor of 10,000. In Fig.~\ref{fig:csr_comp}, we compare the channel switch rate by varying the length of PU duty cycle by a factor of 1/2 and 1/4. For the same PU duty cycle and inter-sensing duration, the channel switch rate increases with decrease in length of PU duty cycle. Also, the channel switch rates are on the order of 500 $\sim$ 2000 switches per second.

The performance metric \textsl{'Channel switch rate'} is closely related to other interesting metrics such as delay in finding the first vacant channel. Note that the secondary transmitter senses a channel after every inter-sensing interval, not after every packet transmission.  Thus, we have carried out MATLAB simulation to plot the delay in finding a first vacant channel against inter-sensing interval in Fig.~\ref{fig:delay} for different PU duty cycle length. We have observed that the delay in finding the first vacant channel increases with increase in inter-sensing duration.

\subsubsection{Energy consumption}
We have also investigated the energy efficiency of the generalized predictive CSA framework. To this end, we first introduce the energy consumption model adopted by our work. The secondary transmitter can be in any one of the following four states during its operation: sensing, transmission, channel switch or back-off (idle) state. The paper \cite{Online} has profiled USRP N210 with WBX daughter board and listed the channel switch energy and delay. However, there is no clear demarcation of the power consumption in the sensing, transmission and idle state. Maleki \textit{et al.} \cite{Maleki2011} have used sensor devices for distributed spectrum sensing in cognitive sensor network. In our work, we set power consumption during sensing, transmission and idle state as 40mW, 16.9mW and 69.5mW, respectively \cite{Maleki2011}. The channel switching energy is set as 20$\mu$J/switch. Further, we have used channel switching delay of 50$\mu$s, which is a realistic assumption.

We have simulated the above energy model in MATLAB by considering 0.8 PU duty cycle on all six channels with HED OFF times (without scaling up parameters) on channels 1, 2, 4, 5 and exponential OFF times on channels 3, 6. The total channel switching energy consumed by secondary transmitter of generalized predictive and predictive CSA frameworks for 10,000 PU ON-OFF cycles are plotted in Fig.~\ref{fig:plot1}. It can be observed that our proposed CSA consumes lesser channel switching energy as compared to predictive CSA in a realistic scenario. We have also verified our results by varying the length of PU duty cycle and inter-sensing interval. In general, the generalized predictive CSA consumes 10\% lesser channel-switching energy than the predictive CSA framework.

The total channel switching energy increases with decrease in inter-sensing interval. However, the throughput of secondary network increases with decrease in inter-sensing interval. Thus, the trade-off between secondary network's throughput and energy consumption motivates one to design optimal inter-sensing interval to maximize the utility which is a function of throughput and energy consumption. The design of optimal inter-sensing interval by itself is a separate topic and have to be studied thoroughly. The preliminary study conducted by us on optimal inter-sensing interval for realistic HED OFF times in single channel scenario indicates that the constant periodic inter-sensing interval is not an optimal policy for distributions other than exponentials.

We have developed generalized predictive CSA framework using realistic PU ON/OFF distributions, and verified its benefits over existing predictive CSA schemes through detailed wireless test-bed and simulation based evaluation studies. Our evaluation results indicate that our proposed generalized CSA framework allows selecting reliable channels that in turn leads to significant gains in terms of throughput and power savings for coexisting networks. 
 
\section{Conclusion}\label{conclusion}
In this paper, we have proposed predictive CSA framework for generalized PU ON-OFF time distribution in multichannel cognitive radio networks. We have approximated the heavy-tailed OFF times of PUs, commonly observed in spectrum measurements, by hyper-exponential distribution and calculated the required probability to use in predictive CSA framework. The performance of generalized predictive CSA framework in terms of the throughput and channel switch rate are evaluated over-the-air by recreating realistic PU occupancy pattern using a wireless test-bed. The detailed simulations have been conducted to study the scalability aspects of the proposed CSA algorithm. The experimental and simulation results suggest considerable reduction in channel switches and energy consumption if ON-OFF time distributions of the PUs are realistically modelled. These results motivate the importance of accurate modelling of ON-OFF time distributions in cognitive radio networks. In our future work, we plan to study the impact of the channel occupancy pattern of PUs on other CRN MAC layer algorithms such as channel sensing interval policy and channel aggregation scheme.

\bibliographystyle{IEEEtran}
\bibliography{library}
\end{document}